\documentclass[11pt,oneside]{article}
\usepackage{setspace,graphicx,amssymb,amsmath,latexsym,amsfonts,amscd,amsthm,multirow,ctable,mathdots,caption,array}
\usepackage{fancyhdr,tabularx,cite,mathrsfs}
\usepackage[headings]{fullpage}
\usepackage{stmaryrd}
\usepackage{rotating}
\usepackage{authblk}
\usepackage{hyperref}

\newcommand{\bea}{\begin{eqnarray}} 
\newcommand{\eea}{\end{eqnarray}} 
\newcommand{\bee}{\begin{eqnarray*}} 
\newcommand{\eee}{\end{eqnarray*}} 
\newcommand{\al}{\begin{align*}} 
\newcommand{\eal}{\end{align*}} 
\newcommand{\be}{\begin{equation}} 
\newcommand{\ee}{\end{equation}} 
\newcommand{\eq}[1]{(\ref{#1})} 
\newcommand{\bem}{\begin{pmatrix}} 
\newcommand{\eem}{\end{pmatrix}} 

\def\a{\alpha} 
\def\b{\beta} 
 
\def\d{\delta}

\def\f{\phi}  
   
\def\h{\eta}

\def\inf{\infty}

\def\l{\lambda} 
\def\m{\mu} 
\def\n{\nu}

\def\p{\pi}    
\def\pa{\partial}

\def\s{\sigma}            
\def\t{\tau} 
\def\th{\theta} 
\def\til{\tilde}

\def\D{\Delta}

\def\L{\Lambda} 
\def\O{\Omega}

\def\Nrt{X}
\def\Srt{\Phi}

\newcolumntype{R}{ >{$}r <{$}}
\newcolumntype{C}{ >{$}c <{$}}
\newcolumntype{L}{ >{$}l <{$}}
\newcolumntype{F}{>{\centering\arraybackslash}m{1.5cm}}

\def\ll{\ell}

\newcommand{\comment}[1]{}

\newcommand{\RR}{{\mathbb R}}
\newcommand{\CC}{{\mathbb C}}
\newcommand{\ZZ}{{\mathbb Z}}
\newcommand{\QQ}{{\mathbb Q}}
\newcommand{\HH}{{\mathbb H}}

\newcommand{\Aut}{\operatorname{Aut}}

\newcommand{\tr}{\operatorname{{tr}}}

\newcommand{\Sym}{{\textsl{Sym}}}

\newcommand{\Dih}{{\textsl{Dih}}}

\newcommand{\ex}{\operatorname{e}} 

\newcommand{\PSL}{\operatorname{\textsl{PSL}}}    
\newcommand{\SL}{\operatorname{\textsl{SL}}}      
\newcommand{\PGL}{\operatorname{\textsl{PGL}}}    
\newcommand{\AGL}{{\textsl{AGL}}}    
\newcommand{\GL}{{\textsl{GL}}}      
\newcommand{\SU}{\operatorname{\textsl{SU}}}    

\newcommand{\G}{\Gamma}	

\newcommand{\rs}{{X}}	

\newcommand{\MM}{\mathbb{M}}	

\theoremstyle{definition}

\theoremstyle{remark}

\numberwithin{equation}{section}

\begin{document}

\setstretch{1.4}

\title{
\vspace{-35pt}
    \textsc{\huge{ Umbral Moonshine and $K3$ Surfaces
    }  }
}

\author[1]{Miranda C. N. Cheng\thanks{mcheng@uva.nl}}
\author[2]{Sarah Harrison\thanks{sarharr@stanford.edu}}

\affil[1]{Institute of Physics and Korteweg-de Vries Institute for Mathematics,\newline
University of Amsterdam, Amsterdam, the Netherlands\footnote{On leave from CNRS, Paris.}}
\affil[2]{Stanford Institute for Theoretical Physics, Department of Physics\\
and Theory Group, SLAC\\
Stanford University, Stanford, CA 94305, USA}
\date{}

\maketitle

\abstract{
Recently, 23 cases of umbral moonshine, relating mock modular forms and finite groups, have been discovered in the context of the 23 even unimodular Niemeier lattices. 
One of the 23 cases in fact coincides with the so-called Mathieu moonshine, discovered in the context of $K3$ non-linear sigma models.
In this paper we establish a uniform relation between all 23 cases of umbral moonshine and $K3$ sigma models, and thereby take a first step in placing umbral moonshine into a geometric and physical context. This is achieved by relating the ADE root systems of the Niemeier lattices to the ADE du Val singularities that a $K3$ surface can develop, and the configuration of smooth rational curves in their resolutions. A geometric interpretation of our results is given in terms of the  marking of $K3$ surfaces by Niemeier lattices. 
}

\clearpage

\tableofcontents

\clearpage

\section{Introduction and Summary} 
\label{Introduction}

Mock modular forms are interesting functions playing an increasingly important role in various areas of mathematics and theoretical physics. 
The ``Mathieu moonshine" phenomenon relating certain mock modular forms and the sporadic group $M_{24}$ was surprising, and its apparent relation to non-linear sigma models of $K3$ surfaces even more so. The fundamental role played by two-dimensional  supersymmetric conformal field theories and $K3$ compactifications makes this moonshine relation interesting not just for mathematicians but also for string theorists. In 2013 it was realised that this Mathieu moonshine is but just one case out of 23 such relations, called  ``umbral moonshine". The 23 cases admit a uniform construction from the 23 even unimodular positive-definite lattices of rank 24 labeled by their non-trivial root systems. While the discovery of these 23 cases of moonshine perhaps adds to the beauty of the Mathieu moonshine relation, it also adds more mystery. In particular, it was previously entirely unclear what the physical or geometrical context for these other 22 instances of umbral moonshine could be. In this paper we establish a relation between  $K3$ sigma models and all 23 cases of umbral moonshine, and thereby take a first step in incorporating umbral moonshine into the realm of geometry and theoretical physics.

\vspace{8pt}
\noindent
\underline{\it Background}

In mathematics, the term ``moonshine" is used to refer to a particular type of relation between modular objects and finite groups. 
It was first introduced to describe the remarkable ``monstrous moonshine" phenomenon  \cite{conway_norton} relating modular functions such as the $J$-function discussed below and the ``{Fischer--Griess monster} group" $\MM$, the largest of the 26 sporadic groups in the classification of finite simple groups. 
The study of this mysterious phenomenon was  initiated by the observation by J. McKay that the second coefficient in the Fourier expansion of the modular function
\begin{align}\label{eqn:intro:FouExpJ}
J(\t)&=J(\t+1)=J(-1/\t) \\ \notag
&=\sum_{m \ge -1} a(m)\, q^{m}= q^{-1} + 196884\, q + 21493760 \,q^2 + 864299970 \,q^3  + \cdots\; \,\,
\end{align}
with $q=e^{2\p i \t} $
satisfies $196884=196883+1$, and  $196883$ is precisely the dimension of the smallest non-trivial representation of $\MM$. 
Note that the $J$-function has the mathematical significance as the unique holomorphic function on the upper-half plane $\HH$ invariant under the natural action of $PSL_2(\ZZ)$ generated by $\t \to \t+1$ and $\t \to -1/\t$, that moreover has the behaviour $J(\t) = q^{-1} + O(q)$ near the cusp $\t\to i\inf$. 
Why and how the specific modular functions and the monster group, usually thought of as belonging to two very different branches of mathematics, are related to each other, remained a puzzle until about a decade after its discovery. 

The key structure that unifies the two turns out to be that of a (chiral) 2d conformal field theory (CFT), or vertex operator algebra in more mathematical terms \cite{FLMPNAS,FLMBerk}. 
The two sides of moonshine -- the modularity and the finite group symmetry -- can naturally be viewed as the manifestation of two kinds of symmetries -- the world-sheet and the space-time symmetries-- the CFT possesses. 
The mathematical proof of monstrous moonshine is achieved by constructing a generalised Kac--Moody algebra based on the above chiral CFT and utilising the no-ghost theorem of string theory, which roughly corresponds  to considering the full 26 dimensions including the 2 light-cone directions of the bosonic string theory  \cite{borcherds_monstrous}. 
We refer to, for instance, \cite{MR2385372} for an introduction on the theory of modular forms and to \cite{gannon} or the introduction of \cite{UM} for a summary of monstrous moonshine.

In 2010, an entirely unexpected new observation, pointing towards a new type of moonshine relating ``mock modular forms" and finite groups, was made in the context of the elliptic genus of $K3$ surfaces. 
Mock modular forms embody a novel variation of the concept of modular forms and are interesting due to their significance in number theory as well as a wide range of applications (cf. \eq{def:shadow}). 
See, for instance,  \cite{Folsom_what,zagier_mock} for an expository account on mock modular forms. 
From a physical point of view, as demonstrated in a series of recent works, the ``mockness" of mock modular forms is often related to the non-compactness of relevant spaces in the theory. See, for instance, \cite{VafaWitten1994,Troost:2010ud,Dabholkar:2012nd,Alexandrov:2012au}. 

As we will discuss in more detail in \S\ref{Singularities K3 Elliptic Genus},  the elliptic genus ${\bf EG}(K3)$ of $K3$ surfaces enumerates the BPS states of a $K3$ non-linear sigma model, and by taking the ${\cal N}=4$ superconformal symmetry of this theory into account, one arrives at a weight 1/2 mock modular form with Fourier expansion \cite{Eguchi1987,Eguchi1988,Eguchi1989}
\begin{align}\label{def:H21}
H^{\Nrt=A_1^{24}}_1 (\t)
 = 2 q^{-1/8} (-1  + 45 \, q+ 231 \,q^2 + 770 \,q^3 + O(q^4)).
\end{align}
The observation by Eguchi--Ooguri--Tachikawa then states that the numbers $45$, $231$, and $770$ are all dimensions of certain irreducible representations of the sporadic Mathieu group $M_{24}$ \cite{Eguchi2010}.
This connection has since been studied, refined, extended, and finally established in \cite{Cheng2010_1,Gaberdiel2010,Gaberdiel2010a,Eguchi2010a,Cheng2011,Gaberdiel:2012gf,2012arXiv1212.0906C,Gaberdiel:2013nya,Persson:2013xpa,Raum,Gannon:2012ck}. 
From a mathematical point of view, the prospect of a novel type of moonshine for mock modular forms is extremely exciting.  
From a physical point of view, the ubiquity of $K3$ surfaces and the importance of BPS spectra in the study of string theory makes this ``Mathieu moonshine" potentially much more relevant than the previous monstrous moonshine. 
See \cite{CheDun_M24MckAutFrms} for a review and 
\cite{Taormina2010,Taormina:2011rr,Gaberdiel2011,Govindarajan:2011em,Cheng:2013kpa},\cite{Taormina:2013jza,Harrison:2013bya,Harvey:2013mda,Wrase:2014fja,Creutzig:2013mqa,John_Sander} for some of the explorations in string theory and $K3$ conformal field theories inspired by this connection.

In 2013, the above relation was realised to be just the tip of the iceberg, or less metaphorically just one case out of a series of such relations, called  ``umbral moonshine" \cite{UM,Cheng:2013wca}. As will be reviewed in more detail in \S\ref{Umbral Moonshine and Niemeier Lattices}, to each one of the 23 Niemeier lattices $L^\Nrt$ -- the 23 even unimodular positive-definite lattices of rank 24 labeled by their non-trivial root systems $\Nrt$ -- one can attach on the one hand a finite group  $G^\Nrt$ and on the other hand a vector-valued mock modular form $H^\Nrt$, such that the Fourier coefficients of $H^\Nrt$ are again suggestive of a relation to certain representations of $G^\Nrt$, analogous to the observation on the functions $J(\t)$ and $H^{\Nrt=A_1^{24}}_1 (\t)$ in \eq{eqn:intro:FouExpJ} and \eq{def:H21}. Further evidence for this relation was provided by relating characters of the same $G^\Nrt$-representations to the Fourier coefficients of other mock modular forms $H^\Nrt_g$, for each conjugacy class $[g]$ of $G^\Nrt$. More precisely, it was conjectured that an infinite-dimensional $G^\Nrt$-module $K^\Nrt$ reproduces the mock modular forms $H^\Nrt_g$ as its graded $g$-characters.
The finite group $G^\Nrt$ is defined by considering the symmetries of the Niemeier lattice $L^\Nrt$, while the mock modular form is determined by its root system $\Nrt$.
The important role played by the rank 24 root systems $\Nrt$ suggests the importance of the corresponding 24-dimensional representation of $G^\Nrt$. 
For instance, for the Niemeier lattice with the simplest root system $\Nrt = A_1^{24}$, the mock modular form  $H^{A_1^{24}}$  is simply given by the function \eq{def:H21} above, and the finite group is $G^X \cong M_{24}$. In this case the umbral moonshine {\it is} the Mathieu moonshine first observed in the context of the $K3$ elliptic genus that we described above. Given the uniform construction of the 23 instances of umbral moonshine from the Niemeier lattices $L^\Nrt$, one is  naturally led to the following questions: {\it What about the other 22 cases of umbral moonshine with $X\neq A_1^{24}$? }
{\it What, if any, is the physical and geometrical relevance of umbral moonshine? }
{\it Are they also related to string or conformal field theories on $K3$? } 
{\it What is the relation between $K3$ and the Niemeier lattices $L^\Nrt$? And the group $G^\Nrt$? The mock modular form $H^\Nrt$ and the underlying $G^\Nrt$--module $K^\Nrt$? }

\vspace{8pt}
\noindent
\underline{\it Summary}

In the present paper we propose a first step in answering the above questions. 
To discuss the relation between  the mock modular form $H^\Nrt$  and the $K3$ elliptic genus, we first take a closer look at the construction of $H^\Nrt$ from the root system $\Nrt$. For any of the 23 Niemeier lattices, the root system is a union of simply-laced root systems with an ADE classification with the same Coxeter number $m$. 
As is well-known, a wide variety of elegant structures in mathematics and physics admit an ADE classification. 
Apart from the simply-laced root systems, another such structure that will be important for us is that of modular invariant combinations of characters of the $A^{(1)}_1$ Kac--Moody algebra at level $m-2$ \cite{Cappelli:1987xt}. 
As  will be reviewed in more detail in \S\ref{The Elliptic Genus of Du Val Singularities}, this classification leads to the introduction of the so-called Cappelli--Itzykson--Zuber matrices for every ADE root system, and these matrices in turn determine the relevant mock modular properties, which uniquely determine $H^\Nrt$ when combined with a certain analyticity condition. 
Hence, the Cappelli--Itzykson--Zuber matrices $\O^\Srt$ constitute a key element in the construction of the 23 instances of umbral moonshine. 

By itself, the question of the classification of certain modular invariants 
 seems remote from any physics or geometry. However, the parafermionic description of the ${\cal N}=2$ minimal models relates this classification to that of the ${\cal N}=2$ minimal superconformal field theories \cite{Cappelli:1986hf,Gepner:1987qi,Vafa:1988uu,Martinec:1988zu}. 
 Moreover,  their seemingly mysterious ADE classification can be related to the ADE classification of du Val (or Kleinian, or rational) surface singularities \cite{Vafa:1988uu,Martinec:1988zu}, whose minimal resolution gives rise to smooth rational (genus 0) curves with intersection given by the corresponding ADE Dynkin diagram. 
 A third way to think about the ADE classification is the fact that these du Val  singularities are isomorphic to the quotient singularity $\CC^2/G$, with $G$ being the finite subgroup of $SU_2(\CC)$ with the corresponding ADE classification \cite{MR0084174}. 
Therefore, a  perhaps simple-minded but logical step towards understanding the physical and geometrical context of umbral moonshine would be  to take the ADE origin of the mock modular form $H^\Nrt$ seriously. In particular we would like to explore if the ADE-ology in umbral moonshine can be related to that of the du Val  singularities.

Recall that the du Val  singularities are precisely the singularities a $K3$ surface can develop. After computing the elliptic genus of du Val  singularities (see \S\ref{The Elliptic Genus of Du Val Singularities}), one realises that the $K3$ elliptic genus can  naturally be split into two parts: one is the contribution from the configuration of the singularities given by $\Nrt$ and the other is the contribution from the mock modular form $H^\Nrt$. Equipped with the mock modular form $H^\Nrt$ for the other 22 Niemeier root systems $\Nrt$ constructed in umbral moonshine, one finds that the same  splitting holds uniformly for all 23 instances of umbral moonshine (cf. \eq{EG_UM_relation1}). Note that this splitting makes no reference to the ${\cal N}=4$ characters, although for the special case $\Nrt=A_1^{24}$ the two considerations render the same result.

While the above fact might be surprising and suggestive, one should  be careful not to claim a strong connection between umbral moonshine and $K3$ string theory too quickly: it's logically possible that the above relation is just a consequence of the fact that the space of the relevant modular objects, the Jacobi forms of weight 0 and index 1 to be more precise, is very constrained and in fact only one-dimensional. See Appendix \ref{app:Proof} for more details. 

To gather more evidence that the umbral moonshine -- a conjecture on the existence of a $G^\Nrt$--module $K^\Nrt$ which (re)produces the mock modular forms $H^\Nrt_g$, $[g]\subset G^\Nrt$ as its graded characters -- and the $K3$ sigma model, one should compare the way $G^\Nrt$ acts on $K^\Nrt$ with the way the BPS spectrum of the $K3$ CFT transforms under its finite group symmetry $G$, when such a non-trivial $G$ exists. 
Let us first focus on the geometric symmetries of $K3$ surfaces (as opposed to ``stringy" CFT symmetries without direct geometric origins). As we will review in more detail in \S\ref{Geometric Interpretation}, thanks to the global Torelli theorem for $K3$, we know that a finite group $G$ is the group of hyper-K\"ahler-preserving symmetries of a certain $K3$ surface $M$ if and only if it acts on the 24-dimensional $K3$ cohomology lattice $H^\ast(M,\ZZ)$ in a certain way. Relating this 24-dimensional representation of $G$ to the natural 24-dimensional representation of $G^\Nrt$ induced from its action on the root system $\Nrt$, this translates into a criterion for a conjugacy class $[g]\subset G^\Nrt$ to arise as a $K3$ symmetries for each of the 23  $G^\Nrt$. 

On the one hand,  umbral moonshine suggests a ``twined" function $Z_g^\Nrt$ for each $[g]\subset G^\Nrt$, where  $Z_g^\Nrt={\bf EG}(K3)$ for the special case that $[g]$ is the identity class (cf. \eq{def:Zg_UM}). In particular, from this consideration we arrive at a conjecture for the elliptic genus of the du Val singularity twined by its symmetries given by the automorphism of the corresponding Dynkin diagram. 
On the other hand, whenever the CFT admits a non-trivial finite automorphism  group $G$, one can compute  the elliptic genus ``twined" by any $g\in G$. These twined elliptic genera ${\bf EG}_g(K3)$ provide information about the Hilbert space as a representation of $G$. 
As a result, for a conjugacy class $[g]\subset G^\Nrt$ arising from $K3$ symmetries, we have two ways to attach a twined function -- $Z_g^\Nrt$ and ${\bf EG}_g(K3)$ -- to such a ``geometric" conjugacy class of $[g]\subset G^\Nrt$. It turns out that they coincide for all the geometric conjugacy classes $[g]$ of any one of the 23 $G^\Nrt$. 
This identity clearly provides non-trivial evidence that all 23 instances of umbral moonshine are related to $K3$ non-linear sigma models.  
 
Recall that in arriving at the above relation we have interpreted the ADE root systems $\Nrt$ as the configuration of rational curves given by the ADE singularities. The above result hence suggests that it might be fruitful to study the symmetries of different $K3$ surfaces with distinct configurations of rational curves in a different framework corresponding to the 23 cases of umbral moonshine. In fact, this has been implemented in a recent analysis of the relation between the $K3$ Picard lattice, $K3$ symplectic automorphisms, and the Niemeier lattices, through a ``marking" of a $K3$ surface $M$ by one of the $L^\Nrt$ such that the Dynkin diagram obtained from the smooth rational curves of $M$ is a sub-diagram of $\Nrt$
 \cite{Nikulin:2011,Nikulin:2014}. As will be discussed in more detail in \S\ref{Geometric Interpretation}, through this marking by the Niemeier lattice $L^\Nrt$, the root system $\Nrt$ obtains the interpretation as the ``enveloping configuration of smooth rational curves" while the finite group $G^\Nrt$ is naturally interpreted  as the ``enveloping symmetry group" of the $K3$ surfaces that can be marked by the given $L^\Nrt$. On the one hand, this provides a geometric interpretation of our results. 
On the other hand, one can view our results as a moonshine manifestation and extension of the geometric analysis in \cite{Nikulin:2011}. 

The organisation of the paper is as follows. 
In \S\ref{The Elliptic Genus of Du Val Singularities} we compute the elliptic genus of the ADE du Val singularities that  $K3$ surfaces can develop. In \S\ref{Umbral Moonshine and Niemeier Lattices} we review the umbral moonshine construction from 23 Niemeier lattices and introduce the necessary ingredients for later calculations. 
Utilising the results of \S\ref{The Elliptic Genus of Du Val Singularities},  in \S\ref{Singularities K3 Elliptic Genus} we establish the relation between the (twined) elliptic genus and the mock modular forms of umbral moonshine. 
In \S\ref{Geometric Interpretation} we provide a geometric interpretation of this result.  
 In \S\ref{Conclusion and Discussion} we close this paper by discussing some open questions and point to some possible future directions. 
In Appendix \ref{sec:modforms} we collect useful definitions. In Appendix \ref{app:Proof} we present the calculations and proofs, and present our conjectures for the twined (or equivariant) elliptic genus for the du Val singularities. The explicit results for the twining functions are recorded in the Appendix \ref{Data}.

\section{The Elliptic Genus of Du Val Singularities}
\label{The Elliptic Genus of Du Val Singularities}
The rational singularities in two (complex) dimensions famously admit an ADE classification. See, for instance, \cite{MR543555}.
They are also called the du Val or Kleinian singularities and are isomorphic to the quotient singularity $\CC^2/G$, with $G$ being the finite subgroup of $SU_2(\CC)$ with the corresponding ADE classification \cite{MR0084174}. 
Any such singularity has a unique minimal resolution. 
The so-called resolution graph, the graph of the intersections of the smooth rational (genus 0) curves of  the minimal resolution, gives precisely the corresponding ADE Dynkin diagram. We will denote by $\Srt$ the corresponding simply-laced irreducible root system. 
In terms of hypersurfaces, it is given by  $W^0_\Srt=0$ with
\begin{align}\label{superpo_ADE_1}
W^0_{A_{m-1}}&= x_1^2 +x_2^2 + x_3^{m}\\
W^0_{D_{m/2+1}}&= x_1^2 +x_2^2x_3 + x_3^{m/2}\\
W^0_{E_{6}}&= x_1^2 +x_2^3 + x_3^{4}\\
W^0_{E_{7}}&= x_1^2 +x_2^3 +x_2 x_3^{3}\\ \label{superpo_ADE_5}
W^0_{E_{8}}&= x_1^2 +x_2^3 + x_3^{5}.
\end{align} 
These singularities show up naturally as singularities of $K3$ surfaces and play an important role in various physical setups, such as in heterotic--type II dualities and in geometric engineering, in  string theory compactifications. See, for instance, \cite{witten_string_1995,Aspinwall:1995zi} and \cite{Katz:1996fh}. 

The 2d conformal field theory description of these (isolated) singularities was proposed in \cite{ooguri_two-dimensional_1995} to be the  product of a non-compact super-coset model $\frac{SL(2,\RR)}{U(1)}$ (the Kazama--Suzuki model \cite{Kazama:1988qp})  and an ${\cal N}=2$ minimal model, followed by an orbifoldisation by the discrete group $\ZZ/m\ZZ$, where $m$ is the Coxeter number of the corresponding simply-laced root system (cf. Table \ref{tab:CoxNum}). In other words, we consider the super-string background that is schematically given by
\be\label{schematic}
{\text{Minkowski space-time}}~\RR^{5,1} \otimes \left( {\cal N}=2~\text{minimal } \otimes  {\cal N}=2~ \frac{SL(2,\RR)}{U(1)}~ {\text{coset}} \right) /({\ZZ/m\ZZ}) . 
\ee
Recall that, when the minimal model is chosen to be the ``diagonal" $A_{m-1}$ theory, the above theory also describes the near-horizon geometry of $m$ NS five-branes \cite{Giveon:1999px}. Note that this point of view plays an important role in the work of \cite{Harvey:2013mda,Harvey:2014cva}, also in the context of discussing the possible physical context of umbral moonshine.

To resolve the singularity let us consider $W^0_\Srt=\m$. In \cite{ooguri_two-dimensional_1995} it was proposed that the sigma model with the non-compact target space $W^0_\Srt=\m$ has an alternative description as the Landau--Ginsburg model with superpotential
\[
\til W_\Srt = -\m x_0^{-m} + W^0_\Srt ,
\]
where $x_0$ is an additional chiral superfield and $m$ is again given by the Coxeter number of $\Srt$. 

The purpose of the rest of the section is to compute the elliptic genus of (the supersymmetric sigma model with the target space being) the du Val singularities. 
First let us focus on the minimal model part. The ${\cal N}=2$ minimal models are known to have an ADE classification \cite{Cappelli:1986hf,Gepner:1987qi,Vafa:1988uu,Martinec:1988zu}
\footnote{Strictly speaking, this classification applies when one requires the presence of a spectral flow symmetry. \\See for instance \cite{MR1449324,MR2925130}  for a discussion on related subtleties.}, based on an ADE classification of the modular invariant combinations of chiral (holomorphic) and anti-chiral (anti-holomorphic) characters of the $A^{(1)}_1$ Kac--Moody algebra \cite{Cappelli:1987xt}. In this language, the ADE classification can be thought of as a classification of the possible  ways to consistently combine left- and right-movers. 
To be more precise, in \cite{Cappelli:1987xt} it was found that a physically acceptable and modular invariant combination of characters of the $A^{(1)}_1$ Kac--Moody algebra at level $m-2$ is necessarily given by a $2m\times 2m$ matrix $\O^\Srt$ corresponding to an ADE root system $\Srt$, 
where we say that a modular invariant is physically acceptable if it satisfies certain integrality, positivity and normalisation conditions. See \cite{Cappelli:1987xt} for more details. 
 The list of these matrices is given in Table \ref{ADE1}. The relation between $\O^\Srt$ and the ADE root system $\Srt$ lies in the following two facts. 
First, $\O^\Srt$ is a $2m\times 2m$ matrix where $m$ is the Coxeter number of $\Srt$. 
Moreover,  $ \O^\Srt_{r,r}-\O^\Srt_{r,-r}=\a_r^\Srt$ for $r=1,\dots,m-1$ coincides with the multiplicity of $r$ as a Coxeter exponent of $\Srt$ (cf. Table \ref{tab:CoxNum}). 
Recall that a Coxeter element $\prod_{i=1}^r r_i$ of the Weyl group of a rank-$r$ root system is the product of reflections with respect to all simple roots (the order in which the product is taken does not change the conjugacy class of the element), and the Coxeter number is the order of such a Coxeter element.

\begin{table}[h]
\captionsetup{font=small}
\centering
\begin{tabular}{c|cccccc}
\toprule
	&$A_{  m-1}$&$D_{1+{  m}/2}$&$E_6$&$E_7$&$E_8$\\
	\midrule
	Coxeter &\multirow{2}*{${  m}$}&\multirow{2}*{${  m}$}&\multirow{2}*{$12$}&\multirow{2}*{$18$}&\multirow{2}*{$30$} \\ 
	number\\\midrule
 Coxeter & \multirow{2}*{$1,2,3,\dots,{  m-1}$} & $1,3,5,\dots,{  m}-1,$ & 1,4,5,& 1,5,7,9,&1,7,11,13,\\
 exponents&&${  m}/{2}$&$7,8,11$&11,13,17 &17,19,23,29 \vspace{1.5pt}\\ 
 \bottomrule
\end{tabular}
\caption{\label{tab:CoxNum}{Simply-laced root systems, Coxeter numbers and Coxeter exponents}}
\end{table}

A quantity that played an important role in the the CFT/LG correspondence \cite{Witten:1993yc} as well as in the recent developments of mock modular form moonshine is the elliptic genus. 
From a physical point of view, the elliptic genus for a 2d ${\cal N}=(2,2)$ superconformal field theory ${\cal T}$ is defined as  \cite{WittenInt.J.Mod.Phys.A9:4783-48001994}  
\be\label{def:elliptic_genus}
Z_{\cal T}(\t,z) = \tr_{{\mathcal H}_{{\cal T},\text {RR}}}\Big( (-1)^{F_R+F_L} y^{J_0} q^{H_L}\bar q^{H_R}  \Big)
\ee
where $F_{R,L}$ denotes the right- and left-moving fermion number respectively. Moreover, the left-  (right-) moving Hamiltonian is given by $H_L= L_0 - c_L/24$ ($H_R= \bar L_0 - c_R/24$ ), where $J_0, L_0,\bar J_0, \bar L_0$ are the zero modes of the left- and right-moving copies of the $U(1)$ R-current and Virasoro parts of the ${\mathcal N}=2$ superconformal algebra, respectively. ${\mathcal H}_{{\cal T},RR}$ denotes the space of quantum states of theory ${\cal T}$ in the Ramond--Ramond sector, and $c_L$ and $c_R$ denote the left- and right-moving central charge of the SCFT. 
 In the above formula, $\t$ takes values in the upper-half plane $\mathbb H$ while $z$ takes values in the complex plane $\CC$, and we have written $q=\ex(\t)$ and $y=\ex(z)$. Throughout the paper we use 
\(
\ex(x) : = e^{2\p i x}.
\)
Because of the insertion $(-1)^{F_R}$, the elliptic genus only receives contributions from left-moving states that are paired with a right-moving Ramond ground state and is therefore holomorphic, at least when the spectrum of the theory is discrete. As such, it is rigid in the sense of being invariant under any continuous deformation of the theory. 

The elliptic genus of the ${\cal N}=2$ minimal model can be computed in various  ways. 
First, from the relation to the parafermion theory, we obtain that the building block of the elliptic genus is the function $\til \chi^r_s (\t,z)$, where $|s| \leq r-1 < m$ \cite{Qiu:1987ux,Gepner:1987qi}.
See Appendix \ref{app:Proof} for the definition of $\til \chi^{r}_{s}$.
From the known spectrum of the minimal model given in terms of the matrix $\O^\Srt$ and the identity $\til \chi^r_s (\t,0) = \d_{r,s}-\d_{r,-s}$ it is straightforward to see that the elliptic genus of the minimal model corresponding to the ADE root system $\Srt$ is given by \cite{kawai_elliptic_1993,DiFrancesco:1993dg}
\begin{align}\label{def:EG_minimal_coset}
Z_{\text{minimal}}^\Srt (\t,z) =  \sum_{r,r' \in \mathbb Z/2m \mathbb Z} \O^\Srt_{r,r'} \til \chi^r_{r'}(\t,z) ={\text{Tr}( \O^\Srt \cdot \til \chi)}   .
\end{align} 
We again refer to Appendix \ref{app:Proof} for more details. 

On the other hand,  the Landau-Ginzburg description facilitates a free-field computation for the elliptic genus  
and  one obtains an infinite-product expression for $Z_{\text{minimal}}^\Srt (\t,z)$ \cite{WittenInt.J.Mod.Phys.A9:4783-48001994}. In terms of the Jacobi theta function \eq{def:JacTheta}, the results are \cite{WittenInt.J.Mod.Phys.A9:4783-48001994,DiFrancesco:1993dg}
  \be\label{minimal_EG_omega}
Z^\Srt_{\rm minimal}=\frac{\theta_1\left(\tau, {m-1\over m}z\right )}{\theta_1\left(\tau, {z\over m}\right )} ~~{\text{for}} ~~\Srt=A_{m-1}
\ee
for the A-series where $m\geq 2,$
\be 
Z^\Srt_{\rm minimal}=\frac{\theta_1\left(\tau, {m-2\over m}z\right )\theta_1\left(\tau, {m+2\over 2m}z\right )}{\theta_1\left(\tau, {2z\over m}\right )\theta_1\left(\tau, {m-2\over 2m}z\right )} ~~{\text{for}} ~~\Srt={D_{{m\over 2}+1}}
\ee
for the D series where $m\geq 6$ and even, and finally 
\begin{align}
Z^{E_6}_{\rm minimal}&=\frac{\theta_1\left(\tau, {3\over 4}z\right )\theta_1\left(\tau, {2\over 3}z\right )}{\theta_1\left(\tau, {z\over 4}\right )\theta_1\left(\tau, {z\over 3}\right )}
\\
Z^{E_7}_{\rm minimal}&=\frac{\theta_1\left(\tau, {7\over 9}z\right )\theta_1\left(\tau, {2\over 3}z\right )}{\theta_1\left(\tau, {2\over 9}z\right )\theta_1\left(\tau, {z\over 3}\right )}
\\
Z^{E_8}_{\rm minimal}&=\frac{\theta_1\left(\tau, {4\over 5}z\right )\theta_1\left(\tau, {2\over 3}z\right )}{\theta_1\left(\tau, {z\over 5}\right )\theta_1\left(\tau, {z\over 3}\right )}
\end{align}
for the $E$-type cases. 
The central charge of these minimal models are given by the Coxeter number $m$ of the corresponding simply-laced root system by
\be 
\hat c= c/3= 1-{2\over m}.
\ee

In order to obtain the elliptic genus of the isolated ADE singularities, another ingredient we need is the elliptic genus of the $\frac{SL(2,\RR)}{U(1)}$ super-coset model. The $\frac{SL(2,\RR)}{U(1)}$ super-coset model is known to describe the geometry of a semi-infinite cigar (a 2d Euclidean black hole) \cite{Witten:1991yr} and is mirror to the ${\cal N}=2$ super Liouville theory \cite{Giveon:1999px,Hori:2001ax}. The level of the super-coset model is related to the mass of the corresponding 2d black hole, and the central charge of the super Liouville theory.  Here, we will consider ${SL(2,\RR)}$ super-current algebra of (super) level $m$. The central charge of the corresponding super-coset theory is 
\[
\hat c = 1+ \frac{2}{m}.
\]
Due to the presence of the adjoint fermions, there is a shift between the level of the super Kac--Moody algebra $[\hat g]_k $ and the level of its bosonic sub-algebra $\hat g_{\bar k}$ given by the corresponding quadratic invariant as
\[
\bar k = k - c_2(g) \quad,
\]
which is given explicitly in terms of structure constants by $c_2(g) \d_{ab}  = f_{a}^{\;\;cd}f_{bcd}$.

The spectrum of the super-coset model and the corresponding torus conformal blocks has been discussed in \cite{Israel:2004ir,Eguchi:2004yi}, following the earlier work  \cite{Dixon:1989cg,Hanany:2002ev}. Since the model is non-compact, the spectrum not surprisingly contains both discrete  and continuous states. In the geometric picture, the discrete states are those localised at the tip of the cigar while the continuous ones are those states whose wave-functions spread into the infinitely long half-cylinder and are only present above a ``mass gap" $\frac{1}{4m}$ on the conformal weight \cite{Dijkgraaf:1991ba}. The fact that the  torus conformal blocks of the super-coset theory coincide with the characters of the corresponding highest weight representations of the ${\cal N}=2$ superconformal algebra constitutes  non-trivial evidence for its equivalence to the ${\cal N}=2$ super Liouville theory. 
Moreover, the continuous states correspond to massive (or long) ${\cal N}=2$ highest weight representations while the discrete states correspond to massless (or short) ones. 
As such, it is easy to see from the Hilbert space (Hamiltonian) definition \eq{def:elliptic_genus} of the elliptic genus that it only receives contribution from the discrete part of the spectrum. 
Accepting the above argument, the building block of the elliptic genus is the Ramond character graded by $(-1)^F$ 
\[
{\rm Ch}^{(\til R)}_{\rm massless}(\t,z;s) =\frac{i \th_1(\t,z)}{\eta^3(\t)} \sum_{k\in \ZZ}y^{2k} q^{mk^2}\frac{(yq^{mk})^{\frac{s-1}{m}}}{1-yq^{mk}}
\]
where $\eta(\t) = q^{1/24}\prod_{n\geq 1}(1-q^n)$ is the Dedekind eta function and $s/2$ is the $U(1)$ charge of the highest weight. The above formula can also be identified as ${\cal N}=2$ characters extended by spectral flow.
Putting them together, from the spectrum of the super-coset model it is straightforward to work out the elliptic genus of the theory
\be
Z_{L_m}(\t,z)=\frac{1}{2} \sum_{s=1}^{m}{\rm Ch}^{(\til R)}_{\text{massless}} (\t,z;m+2-s)+{\rm Ch}^{(\til R)}_{\text{massless}} (\t,z;s)= \frac{1}{2}  \m_{m,0}\big(\t,\frac{z}{m}\big) \frac{i\theta_1(\tau,z)}{\eta(\tau)^3} , 
\ee
where we have used the (specialised) Appell--Lerch sum 
\be\label{def:mu_0}
\m_{m,0}(\t,z) = - \sum_{k\in\mathbb Z} q^{m{k}^2}y^{2km}\frac{1+yq^k }{1-yq^k}  .
\ee
The above partition function has also been calculated in \cite{Troost:2010ud} using an alternative free-field representation of the theory. See also \cite{Eguchi:2010cb,Ashok:2011cy}.

From this we can derive the elliptic genus of the super coset theory coupled to the rational theory 
\[
 \left( {\cal N}=2~\text{minimal } \otimes  {\cal N}=2~ \frac{SL(2,\RR)}{U(1)}~ {\text{coset}} \right) /({\ZZ/m\ZZ}) ,
\]
describing the corresponding du Val surface singularities of type $\Srt$, by using the orbifoldisation formula \cite{kawai_elliptic_1993}
\begin{align}\label{def:EG_ADE}
Z^{\Srt,S}(\t,z) &=  \frac{1}{m} \sum_{a,b \in \ZZ/m\ZZ} q^{a^2} y^{2a} \, Z_{\rm minimal}^\Srt(\t,z+a\t+b) Z_{L_m}(\t,z+a\t+b) \\
&= \frac{1}{2m} \frac{i\th_1(\t,z)}{\eta^3(\t)} \sum_{a,b \in \ZZ/m\ZZ} (-1)^{a+b} q^{a^2/2} y^{a} \, Z_{\rm minimal}^\Srt(\t,z+a\t+b) ~\mu_{m,0}(\t,\frac{z+a\t+b}{m}).
\end{align}

Note that the above elliptic genus is not modular, as opposed to the familiar situation with elliptic genera of a supersymmetric conformal field theory. 
In fact, it is mock modular in the following sense\cite{Zwegers2008}. Let the ``completion" of $ \m_{m,0} (\t,z)$ be
\be\label{pole_completion}
\hat \m_{m,0} (\t,z) =  \m_{m,0} (\t,z) -\ex(-\tfrac{1}{8}) \,\frac{1}{\sqrt{2m}}  \sum_{r \in \ZZ/2m\ZZ} \th_{m,r}(\t,z)  \int^{i\inf}_{-\bar \t}  (\t'+\t)^{-1/2} \overline{S_{m,r}(-\bar \t')} \, {\rm d}\t', 
\ee
then $\hat \m_{m,0}$ transforms like a Jacobi form of weight $1$ and index $m$ under the Jacobi group $\SL_2(\ZZ)\ltimes \ZZ^2$ but is not holomorphic. (See Appendix \ref{sec:modforms} for the definition of Jacobi forms.)
In the above formula,  
$S_m = (S_{m,r})$ denotes the vector-valued cusp form for $\SL_2(\ZZ)$ whose components are given by the unary theta function (cf. \eq{def:theta})
\be\label{def:unaryS}
S_{m,r}(\t) = \sum_{ k = r\!\! \pmod{2m}} \, k \, q^{k^2/4m}   = \frac{1}{2\p i}\frac{\pa}{\pa z} \th_{m,r}(\t,z)\lvert_{z=0}.
\ee

In Appendix \ref{sec:compute_twining}, we will also conjecture the answer for the elliptic genera of these ADE-singularities twined by automorphisms of the corresponding Dynkin diagram, which can be thought of as permuting the smooth rational curves in the minimal resolution.  

This absence of the usual modularity can be attributed to the fact that the target space of the theory is non-compact and hence the spectrum contains a continuous part \cite{Troost:2010ud}. This is however seemingly in contradiction with the expectation that a path integral formulation of the elliptic genus should render a function transforming nicely under $SL_2(\ZZ)$, corresponding to the $SL_2(\ZZ)$ mapping class group of the world-sheet torus underlying the path integral formulation. This issue has been recently addressed in \cite{Troost:2010ud}, and further refined in \cite{Ashok:2013pya,Murthy:2013mya}, for the cigar theory. These authors found that a path integral computation indeed renders an answer that is modular but non-holomorphic, and the breakdown of holomorphicity is attributed to the imperfect cancellation between contributions of the bosonic and fermonic states to the elliptic genus \eq{def:elliptic_genus} in the continuous part of the spectrum. 
Analogously, we expect the path integral formulation of the elliptic genus of the ADE singularities will render as the answer the real Jacobi form
\begin{align}
\hat Z^{\Srt,S}(\t,z) &= \frac{1}{2m} \frac{i\th_1(\t,z)}{\eta^3(\t)} \sum_{a,b \in \ZZ/m\ZZ} (-1)^{a+b} q^{a^2/2} y^{a} \, Z_{\rm minimal}^\Srt(\t,z+a\t+b) ~\hat \mu_{m,0}(\t,\frac{z+a\t+b}{m}).
\end{align}

Finally, we note that there is a different  definition of elliptic genus that is purely geometric.
For a compact complex manifold $M$ with dim$ _\CC M=d_0$, the elliptic genus is defined as the character-valued Euler characteristic of the formal vector bundle 
\cite{Ochanine,Witten1987,Landweber_book,WittenInt.J.Mod.Phys.A9:4783-48001994,KawaiNucl.Phys.B414:191-2121994}
$$
{\bf E}_{q,y} 
=
y^{d/2}{\textstyle \bigwedge}{ }_{-y^{-1}} T_M^\ast
\textstyle{\bigotimes}_{n\geq 1} \textstyle\bigwedge{ }_{-y^{-1}q^n} T_M^\ast\bigotimes_{n\geq 1} \textstyle\bigwedge{ }_{-yq^n} 
T_M \bigotimes_{n\geq 0} S_{q^n} (T_M\oplus  T_M^\ast),
$$
where $T_M$ and \(T_M^\ast\) are the holomorphic tangent bundle and its dual, and we adopt the notation
\[ 
\textstyle\bigwedge{ }_q V = 1 + q V + q^2 \textstyle \bigwedge^2 V + \dots,\quad S_q V = 1 + q V +q^2 S^2V+ \cdots\cdots,
\]
with $S^kV$ denoting the $k$-th symmetric power of $V$. In other words, we have 
\be\label{def_eg}
{\bf EG}(\t,z;M) = \int_M ch({\bf E}_{q,y}) {\rm Td}(M)
\ee
where $ {\rm Td}(M)$ is the Todd class of $T_M$. 
For $M$ a (compact) Calabi--Yau manifold, the above geometric definition and the conformal field theory definition, when the CFT is taken to be the 2d non-linear sigma model of $M$, are believed to give the same function \cite{Witten1987,Kapustin:2005pt}. 
The fact that the CFT elliptic genus is rigid corresponds to the geometric fact that ${\bf EG}(\t,z;M)$ is a topological invariant. 
Note that the above definition is manifestly holomorphic. We expect that a suitable generalisation of the above definition which handles non-compact geometries will lead to the geometric elliptic genus ${\bf EG}(\t,z;\Srt)=Z^{\Srt,S}(\t,z)$ of the du Val singularity. In this paper we will simply refer to $Z^{\Srt,S}(\t,z)$  as the elliptic genus of the ADE singularity of type $\Srt$.

\section{Umbral Moonshine and Niemeier Lattices} 
\label{Umbral Moonshine and Niemeier Lattices}

In this section we will briefly review the umbral moonshine conjecture and its construction from the 23 Niemeier lattices \cite{Cheng:2013wca}. 
The readers are referred to \cite{Cheng:2013wca} for more details. 
Let us start by recalling what the Niemeier lattices are. 
Consider positive-definite lattices of rank $24$, we would like to know which of them are even and unimodular. 
In string theory, one is often interested in even, unimodular lattices due to the modular invariance of their theta functions. 
In the classification of  positive-definite even unimodular lattices,  a special role will be played by the root system of the lattice $L$, given by  $\Delta(L)= \{v\in L\lvert \langle v, v\rangle =2\}.$  

The even unimodular  positive-definite lattices of rank $24$ were classified by Niemeier \cite{Nie_DefQdtFrm24}.  There are 24 of them (up to isomorphisms). 
The {Leech lattice} is the unique even, unimodular,  positive-definite lattice of rank $24$ with no roots \cite{Con_ChrLeeLat}, discovered shortly before the classification of Niemeier \cite{Lee_SphPkgs,Lee_SphPkgHgrSpc}. 
Apart from the Leech lattice, there are  $23$ other inequivalent even unimodular lattices of rank 24. 
They are uniquely determined by their root systems $\D(L)$, that are all unions of the simply-laced root systems. 
Moreover, the 23 root systems of the 23 Niemeier lattices are precisely the 23 unions of ADE root systems  satisfying the following two simple conditions: first, all of the irreducible components have the same Coxeter numbers; second, the total rank is 24.  They are  listed in Table \ref{tab:mugs}, where $n$ denotes $\ZZ/n\ZZ$. 
Here and in the rest of the paper we will  adopt the shorthand notation $A_{m-1 }^{d_A}  D_{m/2+1}^{d_D} (E^{(m)})^{d_E}$ for the direct sum of 
$d_A$ copies of $A_{m-1 }$, $d_D$ copies of $D_{m/2+1}$ and ${d_E}$ copies of 
\be\label{def:Em}
E^{(m)}=\begin{cases} E_{6}, E_7, E_8 \;&{\rm for } \; m=12,18,30 \\  \emptyset & {\rm otherwise}\end{cases}.
\ee

Let $\Nrt$ be one of the 23 root systems listed above, and denote by $L^\Nrt$ the unique (up to isomorphism) Niemeier lattice with root system $\Nrt$. 
For each of these 23 $L^\Nrt$ we will have an instance of umbral moonshine as we will explain now. 
First, we need to define the finite group relevant for this new type of moonshine. 
Let us consider the automorphism group ${\rm Aut}(L^\Nrt)$ of the lattice $L^\Nrt$. Clearly, any element of the Weyl group ${\rm Weyl}(\Nrt)$ generated by reflections with respect to any root vector leaves the lattice invariant. In fact, ${\rm Weyl}(\Nrt)$ is a normal subgroup of ${\rm Aut}(L^\Nrt)$ and we define the ``{umbral group}" $G^\Nrt$ to be the corresponding quotient
\be\label{def:umbral group}
G^\Nrt = {\rm Aut}(L^\Nrt)/{\rm Weyl}(\Nrt). 
\ee
The list of the 23 $G^\Nrt$ is given in Table \ref{tab:mugs}. 

\begin{table}[h]
\captionsetup{font=small}
\begin{center}
\caption{Umbral Groups}\label{tab:mugs}
\medskip

\begin{tabular}{ccccccccccc}
\multicolumn{1}{c|}{$\rs$}&$A_1^{24}$&$A_2^{12}$&$A_3^8$&$A_4^6$&$A_5^4D_4$&$A_6^4$&$A_7^2D_5^2$\\
	\cmidrule{1-8}
\multicolumn{1}{c|}{$G^{\rs}$}&			$M_{24}$&	$2.M_{12}$&	$2.\AGL_3(2)$&	$\GL_2(5)/2$&	$\GL_2(3)$&	$\SL_2(3)$&$\Dih_4$\\
\multicolumn{1}{c|}{$\bar{G}^{\rs}$}&		$M_{24}$&	$M_{12}$&	$\AGL_3(2)$&		$\PGL_2(5)$&	$\PGL_2(3)$&	$\PSL_2(3)$&$2^2$\\
\\
\multicolumn{1}{c|}{$\rs$}&$A_8^3$&$A_9^2D_6$&$A_{11}D_7E_6$&$A_{12}^2$&$A_{15}D_9$&$A_{17}E_7$&$A_{24}$\\
	\cmidrule{1-8}
\multicolumn{1}{c|}{$G^{\rs}$}&$\Dih_6$&$4$&			$2$&	$4$& $2$&$2$&$2$\\
\multicolumn{1}{c|}{$\bar{G}^{\rs}$}&$\Sym_3$&$2$&		$1$&$2$& $1$&$1$&$1$\\
\\
\multicolumn{1}{c|}{$\rs$}&$D_4^{6}$&$D_6^{4}$&$D_8^3$&$D_{10}E_7^2$&$D_{12}^2$&$D_{16}E_8$&$D_{24}$\\
	\cmidrule{1-8}
\multicolumn{1}{c|}{$G^{\rs}$}&			$3.\Sym_6$&	$\Sym_4$&	$\Sym_3$&	$2$&	$2$&	$1$&$1$\\
\multicolumn{1}{c|}{$\bar{G}^{\rs}$}&		$\Sym_6$&	$\Sym_4$&	$\Sym_3$&	$2$&	$2$&	$1$&$1$\\
\\
\multicolumn{1}{c|}{$\rs$}&$E_6^4$&$E_8^3$\\
	\cmidrule{1-3}
\multicolumn{1}{c|}{$G^{\rs}$}&	$\GL_2(3)$&$\Sym_3$&\\
\multicolumn{1}{c|}{$\bar{G}^{\rs}$}&	$\PGL_2(3)$&$\Sym_3$&
\end{tabular}
\end{center}
\end{table}

After defining the relevant finite group $G^\Nrt$,  we will now define the relevant (vector-valued) mock modular forms $H^\Nrt_g$, $[g]\subset G^\Nrt$, for the umbral moonshine. As explained in \S\ref{The Elliptic Genus of Du Val Singularities}, the ADE classification of the modular invariant combinations of $\hat A_1^{(1)}$ characters is given by a symmetric matrix $\O^{\Srt}$ of size $2m$, where $m$ denotes the Coxeter number of $\Srt$, for every simply-laced root system $\Srt$. As we have seen, the Cappelli--Itzykson--Zuber matrix $\O^{\Srt}$ also controls the spectrum and hence the elliptic genus \eq{minimal_EG_omega} of the 2d minimal model of type $\Srt$. 
Now consider any one of the 23 Niemeier root systems $\Nrt$ listed above. 
Since they are unions $\Nrt = \cup_i \Srt_i$  of simply-laced root systems $\Srt_i$ with the same Coxeter number, we can extend the definition of the $\O$-matrix to $\O^\Nrt = \sum_i \O^{\Srt_i}$. Using these $\O$-matrices we can then define for each Niemeier lattice $L^\Nrt$ the vector-valued weight 3/2 cusp form 
$$S^\Nrt = \O^\Nrt S_m = (S^\Nrt_r), \quad r\in \ZZ/2m\ZZ$$ 
with the $r$-th component given by 
\[
S^\Nrt_r = \sum_{r'\in \ZZ/2m\ZZ} \O^\Nrt_{r,r'} S_{m,r'}
\]
in terms of the unary theta function \eq{def:unaryS}. 
 From $ (\O^\Nrt)_{r,r'}= (\O^\Nrt)_{-r,-r'}$ and $S_{m,r} =-S_{m,-r}$ it is easy to see that $S^\Nrt_r = -S^\Nrt_{-r}$. 
 
Given the cusp form $S^\Nrt$, we can now specify the mock modular form $H^\Nrt$ by the following two conditions. 
First we specify its mock modular property:  we require $H^\Nrt$ to be a weight 1/2 vector-valued mock modular form whose shadow is given by $S^\Nrt$. 
More precisely, let
\[
\hat H^\Nrt_r (\t)= H^\Nrt_r(\t) +\ex(-\tfrac{1}{8})  \,\frac{1}{\sqrt{2m}}   \int^{i\inf}_{-\bar \t}  (\t'+\t)^{-\frac{1}{2}} \,\overline{S^\Nrt_{r}(-\bar \t')} \, {\rm d}\t', 
\]
then 
\[
\sum_{r\in \ZZ/2m\ZZ} \hat H^\Nrt_r(\t)\, \th_{m,r}(\t,z)
\]
 transforms as a Jacobi form of weight $1$ and index $m$ under the Jacobi group $\SL_2(\ZZ)\ltimes \ZZ^2$.  
 Recall that the shadow $s(\t)$ of a mock modular form $f(\t)$ of weight $w$ is the function, a modular form of weight $2-w$ itself for the same $\G<SL_2(\RR)$, whose integral gives the non-holomorphic completion 
 \be\label{def:shadow}
 \hat f(\t)  = f(\t) + \ex(\frac{w-1}{4})\int_{-\bar \t}^{i\inf} (\t+\t')^{-w} \, \overline{s(-\bar \t')}\,d\t'
 \ee
 of $f$ which transforms as a weight $w$ modular form. This definition has a straightforward generalisation to the vector-valued case which we have employed above. 
 
 After specifying the mock modularity, we impose the following analyticity condition : we require 
its growth near the cusp to be
 \be
 q^{1/4m} H^\Nrt_r(\t) = O(1)  \quad {\rm as}\; \t \to i \inf 
 \ee
for every element $r\in \ZZ/2m\ZZ$.   The above two conditions turn out to be sufficient to determine $H^\Nrt$ uniquely (up to a rescaling),  as shown in \cite{Sko_Thesis,Dabholkar:2012nd,Cheng:2013wca}. We also fix the scaling by requiring $ q^{1/4m} H^\Nrt_1(\t) = -2 + O(q).$

For instance, when considering the Niemeier lattice with the simplest root system, $\Nrt=A_1^{24}$, the unique mock modular form determined by the above condition reads
\begin{align}\label{def:H2}
H^{\Nrt=A_1^{24}}_1 (\t)&= -H^{\Nrt=A_1^{24}}_{-1} (\t) = \frac{-2 E_2(\t) + 48 F_2^{(2)}(\t)}{\h(\t)^3}\\
& = 2 q^{-1/8} (-1  + 45 \, q+ 231 \,q^2 + 770 \,q^3 + O(q^4)).
\end{align}
where $E_2(\t)$ stands for the weight 2 Eisenstein series and 
$$
F_2^{(2)}(\t)= \sum_{\substack{r>s>0\\ r-s=1\, {\rm mod }\; 2 }} (-1)^{r} \,s \,q^{rs/2} = q+q^2-q^3+q^4+\dots \;.
$$
As mentioned in \S\ref{Introduction}, 
the first observation that led to the recent development in the moonshine phenomenon for mock modular forms is the fact that the above numbers $45$, $231$, $770$ coincide with the dimensions of certain irreducible representations of the corresponding umbral group $G^{\Nrt}\cong M_{24}$ for $\Nrt=A_1^{24}$.

Note that without the non-holomorphic completion, the function 
$$\sum_{r\in \ZZ/2m\ZZ} H^\Nrt_r\, \th_{m,r}$$
does not transform nicely under the modular group; it is a mock Jacobi form according to the definition given in \cite{Dabholkar:2012nd}. In \cite{Cheng:2013wca}, following \cite{Dabholkar:2012nd},  this mock  Jacobi form is interpreted as the finite part of a meromorphic (as a function of $z$ 
 ) Jacobi form 
 with simple poles at $m$-torsion points.
For later convenience, we will define another mock Jacobi form
\be\label{def:phi}
\f^\Nrt(\t,z) = \frac{i \th_1(\t,mz) \th_1(\t,(m-1)z) }{\eta^3(\t)\th_1(\t,z) }\sum_{r\in \ZZ/2m\ZZ} H^\Nrt_r(\t)\, \th_{m,r}(\t,z)
\ee
which contains exactly the  same information as the vector-valued mock modular form $H^\Nrt$.

In order to relate such functions to representations of the finite group $G^\Nrt$ that we have constructed, we need as many vector-valued functions similar to $H^\Nrt$ as the number of conjugacy classes of $G^\Nrt$ to encode the characters of the underlying representation. 
Hence,  for every Niemeier lattice $\Nrt$, and for every conjugacy class $[g] \subset G^\Nrt$ we would like to define a vector-valued mock modular form $H^\Nrt_g$. As before, first we need to specify their mock modular properties. The relevant congruence subgroup $\Gamma_0(n_g) \subseteq SL_2(\ZZ)$ (see \eq{def:hecke_congruence}), is determined by $n_g$, the order group element $g$.
This is similar to the situation both in monstrous moonshine \cite{conway_norton} and, not unrelatedly, 2-dimensional CFT.

We need two more pieces of data to completely specify the mock modularity of $H^\Nrt_g$. 
The first one is the shadow. 
By studying the action of $\langle g \rangle$, the cyclic group generated by $g$, we can analogously define a $2m\times 2m$ matrix $\O^\Nrt_g$ and the corresponding cusp form $S^\Nrt_g =\O^\Nrt_g S_m =S^\Nrt_{g,r}$. See \S 5.1 of \cite{Cheng:2013wca} for the list of $\O^\Nrt_g$. 
The second piece of data we need is the multiplier system system on $\G_0(n_g)$, namely a projective representation $\n_g : \G_0(n_g) \to GL_{2m}(\CC)$ of the congruence subgroup $\G_0(n_g)$. 
In the case where the specified shadow $S^\Nrt_g$ does not vanish, the definition of the shadow stipulates the multiplier of the mock modular form to be the inverse of the shadow. As a result, this second piece of data is implied by the first. 
If however $S^\Nrt_g=0$, namely when the mock modular form $H_g^\Nrt$ is in fact modular, one needs to specify the multiplier system independently.  
It turns out that  $\n_g$ is identical to the inverse of the multiplier of $S^\Nrt$ on a group $\G_0(n_g h_g)< \G_0(n_g)$ for certain integral $h_g>1$. 
See \cite{Cheng:2013wca} for more details. 
In particular, let 
\[
\hat H^\Nrt_{g,r} (\t)= H^\Nrt_{g,r}(\t) +\ex(-\tfrac{1}{8})  \,\frac{1}{\sqrt{2m}}   \int^{i\inf}_{-\bar \t}  (\t'+\t)^{-1/2} \overline{S^\Nrt_{g,r}(-\bar \t')} \, {\rm d}\t', 
\]
then 
\[
\sum_{r\in \ZZ/2m\ZZ} \hat H^\Nrt_{g,r}(\t)\, \th_{m,r}(\t,z)
\]
transforms like a Jacobi form of weight $1$ and index $m$ under the  group $\Gamma_0(n_gh_g)\ltimes \ZZ^2$. By the same token, the function $\sum_{r\in \ZZ/2m\ZZ} H^\Nrt_{g,r}\, \th_{m,r}$ is a mock Jacobi form of weight $1$ and index $m$ under  $\Gamma_0(n_gh_g)\ltimes \ZZ^2$. 

As before, after specifying the mock modular property we also need to fix the analyticity property of $H^\Nrt_g$. 
For $\G_0(n_g)$ with $n_g>1$, there is more than one cusp (representative), namely more than one $\G_0(n_g)$-orbit among $\QQ\cup i\inf$. 
For the cusp (representative) located at $\t\to i \inf$ we require the same growth condition
 \be
 q^{1/4m} H^\Nrt_{g,r}(\t) = O(1)  \quad {\rm as}\; \t \to i \inf .
 \ee
 for every $ r\in \ZZ/2m\ZZ$. 
 Moreover we require the function to be bounded
  \be
H^\Nrt_{g,r}(\t) = O(1)   \quad {\rm as}\; \t \to \a\in \QQ,  \quad \a \not\in \G_0(n_g) \inf.
 \ee
at all other cusps.

  After specifying the shadow $S^\Nrt_g$, the multiplier system $\n_g$ and the behaviour at the cusps, a vector-valued mock modular form $H_g^\Nrt$ of weight 1/2 for $\G_0(n_g)$ was then given in \cite{Cheng:2013wca} for every $[g]\subset G^\Nrt$ and for all 23 Niemeier lattices $L^\Nrt$.
 See  \cite{Cheng:2013wca} for explicit Fourier coefficients of the $q$-expansions of  $H^\Nrt_{g,r}$. 
  Finally, it was conjectured in  \cite{Cheng:2013wca} that $H_{g}^\Nrt$ is the unique (up to rescaling) vector-valued mock modular form with the above mock modularity and poles.  
For later convenience, we will also define
\be\label{def:phi_g}
\f^\Nrt_g(\t,z) = \frac{i \th_1(\t,mz) \th_1(\t,(m-1)z) }{\eta^3(\t)\th_1(\t,z) }\sum_{r\in \ZZ/2m\ZZ} H^\Nrt_{g,r}(\t)\, \th_{m,r}(\t,z)
\ee
Note that we recover $H^\Nrt$ and $\f^\Nrt$ by putting $[g]$ to be the identity class in the above discussions on $H^\Nrt_g$ and  $\f^\Nrt_g$.

After constructing the finite group $G^\Nrt$ and the set of vector-valued mock modular forms $H^\Nrt_g=(H^\Nrt_{g,r})$ for each Niemeier lattice $L^\Nrt$, we can now formulate the umbral moonshine conjecture  \cite{Cheng:2013wca}. This conjecture states that for  every Niemeier lattice $\Nrt$, for every $1\leq r \leq m-1$ we have an infinite-dimensional $\ZZ$-graded module $K^\Nrt_r = \oplus_{D} K_{r,D}^\Nrt $ for $G^\Nrt$ such that $H^\Nrt_{g,r}$ is essentially given by the graded characters $\sum_{D = -r^2\! \pmod{4m}, D>0} q^{D/4m} {\rm Tr}_{K_{r,D}^\Nrt}   g$, up to the possible inclusion of a polar term $-2 q^{-1/4m}$ and a constant term (as well as an additional factor of 3 in the case $\Nrt=A_8^3$). See \S 6.1 of  \cite{Cheng:2013wca} for the precise statement of the conjecture.
In summary, umbral moonshine conjectures for each of the 23 Niemeier lattices the existence of a special module $K^\Nrt$ of the finite group $G^\Nrt$, which underlies the special mock Jacobi forms $\f_g^\Nrt$. This conjecture has so far  been proven for the case $\Nrt=A_1^{24}$ \cite{Gannon:2012ck}, and explicitly verified till the first hundred terms in the $q$-expansion for the other 22 cases. 
In the following section we will demonstrate the relation between the mock Jacobi forms $\f^\Nrt_g$and the elliptic genus of $K3$ surfaces. Subsequently we will explore the relation between the Niemeier lattices $L^\Nrt$, the  finite group $G^\Nrt$, the (conjectural) $G^\Nrt$-module $K^\Nrt$, and the (stringy) symmetry of $K3$ surfaces.

\section{Umbral Moonshine and the (Twined) $K3$ Elliptic Genus} 
\label{Singularities K3 Elliptic Genus}

In \S\ref{The Elliptic Genus of Du Val Singularities} we have computed the elliptic genus of Du Val singularities a $K3$ surface can develop. 
In \S\ref{Umbral Moonshine and Niemeier Lattices} we have briefly reviewed the umbral moonshine conjecture relating a finite group $G^\Nrt$ and a set of mock Jacobi forms $\f_g^\Nrt$ for every Niemeier lattice $L^\Nrt$ via an underlying $G^\Nrt$-module $K^\Nrt$. 
In this section we will see how these two separate topics meet in the framework of (twined) elliptic genera for $K3$ surfaces.

Let's start by briefly reviewing the relation between the elliptic genus of $K3$ surfaces and the Mathieu group $M_{24}$, which is also the umbral group $G^\Nrt$ for the Niemeier lattice with root system $\Nrt=A_1^{24}$. 
The 2d non-linear sigma model of a $K3$ surface is a CFT with central charge $c=6$ and with  a (small)  ${\cal N}=4$ superconformal symmetry. As explained in \S\ref{The Elliptic Genus of Du Val Singularities}, the elliptic genus \eq{def:elliptic_genus}  is the same for different  $K3$ sigma models and coincides with the geometric elliptic genus of $K3$. It is computed to be (cf. \eq{def:JacTheta}) \cite{Eguchi1989}
\be\label{EG_K31}
 {\bf EG}(\t,z;K3) 
	= 8 \sum_{i=2,3,4}\left(\frac{\th_i(\t,z)}{\th_i(\t,0)}\right)^2 . 
\ee
The ${\cal N}=4$ superconformal symmetry of the theory implies that the spectrum is composed of irreducible representations (``multiplets") of the  ${\cal N}=4$ superconformal  algebra, and the elliptic genus permits a decomposition into their characters.  

Recall  that the  ${\cal N}=4$ superconformal algebra contains subalgebras isomorphic to the affine Lie algebra $\hat{\mathfrak{sl}}_2$ and the Virasoro algebra, and in a unitary representation the former of these acts with level $m-1$ and the latter with central charge $c=6(m-1)$ for some integer $m>1$.  The unitary irreducible highest weight representations $v_{m;h,j}$ are labelled by the two {quantum numbers} $h$ and $j$ which are the eigenvalues of $L_0$ and $\frac{1}{2}J_0^3$ of the highest weight state, respectively, when acting on the highest weight state \cite{Eguchi1988,Eguchi1988a}. 
 (We adopt a normalisation of the {$\SU(2)$ current} $J^3$ such that the zero mode $J^3_0$ has integer eigenvalues. The  shift by $-1$ in the central charge and the level of the current algebra is due to the $-1$ difference between the level and the index of the theta functions underlying the characters, as we will see below.) 
 The algebra has two types of highest weight representations: the {\em short} (or {\em BPS}, {\em supersymmetric}) ones and the  {\em long} (or {\em non-BPS}, {\em non-supersymmetric}) ones. 
  In the Ramond sector, the former has $h=\frac{c}{24}=\frac{m-1}{4}$ and $j\in\{ 0,\frac{1}{2},\cdots,\frac{m-1}{2}\}$, while the latter has $h > \frac{m-1}{4}$ and $j\in\{ \frac{1}{2},1,\cdots, \frac{m-1}{2}\}$.
 Their (Ramond) graded characters, defined as 
\be
{\rm ch}_{m;h,j}(\t,z) = \tr_{v_{m;h,j}} \left( (-1)^{J_0^3}y^{J_0^3} q^{L_0-c/24}\right),
\ee
are given by 
\be \label{masslesschar}
	{\rm ch}_{m;h,j}(\t,z)= 
	\frac{i\,\th_1(\t,z)^2}{\eta^3(\t) \th_1(\t,2z)}  \m_{m,j}  (\t,z)
\ee
and 
\be \label{massivechar}
{\rm ch}_{m;h,j}(\t,z) =
\frac{i\,\th_1(\t,z)^2}{\eta^3(\t) \th_1(\t,2z)}  \,q^{h-\frac{c}{24}-\frac{j^2}{m}} \,  \big(\th_{m,2j} (\t,z)-\th_{m,-2j} (\t,z)\big)
\ee
in the short and long cases, respectively \cite{Eguchi1988}. In the above formulas, $\m_{m,j}$ is given by $\m_{m,0}$ \eq{def:mu_0} and the identity
\[
 \m_{m,\frac{r}{2}} =(-1)^{r}  (r+1) \m_{m,0}+ (-1)^{r-n+1}  \sum_{n=1}^r n\, q^{-\frac{(r-n+1)^2}{4m}}  ( \th_{m,r-n+1}-\th_{m,-(r-n+1)})   .
\]

When applying the above formula to the $K3$ sigma models which have  $c=6 \,(m=2)$, we obtain the following rewriting of the function in (\ref{EG_K31}): 
\begin{align}
{\bf EG}(\t,z;K3) &= 20\, {\rm ch}_{2;\frac{1}{4},0}-2 \,{\rm ch}_{2;\frac{1}{4},\frac{1}{2}} + \big( 90\, {\rm ch}_{2;\frac{5}{4},\frac{1}{2}} + 462 \,{\rm ch}_{2;\frac{9}{4},\frac{1}{2}} + 1540\,{\rm ch}_{2;\frac{13}{4},\frac{1}{2}} + \dots \big) \\ \notag
& =\frac{i\,\th_1(\t,z)^2}{\eta^3(\t) \th_1(\t,2z)}  \Big\{ 24 \,\m_{2,0}(\t,z)+ (\th_{2,-1}(\t,z)-\th_{2,1}(\t,z)) \\ \label{EG_K3_decomposition}
&\quad \times  (-2q^{-1/8} + 90 q^{7/8} + 462 q^{15/8}+1540 q^{23/8} 
+ \dots)  \Big\}
\end{align}
where $\dots$ corresponds to terms in ${\bf EG}(\t,z;K3)$ of the form $\frac{i\,\th_1(\t,z)^2}{\eta^3(\t) \th_1(\t,2z)} q^\a y^\b$ with $\a-\b^2/8>3$. 
Note that the $q$-series in the last line is nothing but the umbral mock modular form \eq{def:H2} coresponding to the Niemeier lattice with root system $\Nrt=A_1^{24}$ that we introduced in the previous section. As mentioned in \S\ref{Introduction}, it was precisely in this context of decomposing the $K3$ elliptic genus into ${\cal N}=4$ characters that the first case of moonshine for mock modular forms was observed \cite{Eguchi2010}. 

From the above discussion, we see that the two contributions to ${\bf EG}(\t,z;K3)$, given by
\[
 24 \,\m_{2,0}(\t,z)
\]
and 
\[
 -\sum_{r\in \ZZ/4\ZZ} H^{\Nrt=A_1^{24}}_r(\t)\th_{2,r}(\t,z)   ,
\]
in the $\{\}$ bracket, can roughly be thought of as the contributions from the BPS and non-BPS ${\cal N}=4$ multiplets respectively\footnote{Strictly speaking, the polar term ``$-2q^{-1/8}$" of $H^{\Nrt=A_1^{24}}_1$ also corresponds to the contributions from BPS multiplets, while all the infinitely many other terms are contributions from non-BPS multiplets.}. 

However, there is a possible alternative interpretation, thanks to the identity between the short ${\cal N}=4$ characters and the elliptic genus of an $\Srt=A_1$ singularity:
\begin{align}
Z^{A_1,S} (\t,z) &=  {\rm ch}_{2;\frac{1}{4},0}(\t,z) ,
\end{align} 
which follows from the identity 
\[
\frac{1}{2}\sum_{a,b=0}^1 q^{a^2 } y^{2a} \th_1(\t,z+a\t+b)\,\m_{2,0}(\t,\frac{z+a\t+b}{2}) 
= \frac{\th_1(\t,z)^2}{i\th_1(\t,2z)} \,\m_{2,0}(\t,z) 
.
\]

In other words, we can re-express the elliptic genus of $K3$ as
\begin{align}
{\bf EG}(\t,z;K3) = 24 Z^{A_1,S} (\t,z) - \frac{i\,\th_1(\t,z)^2}{\eta^3(\t) \th_1(\t,2z)}\, \sum_{r\in \ZZ/4\ZZ} H^{A_1^{24}}_r \th_{2,r} (\t,z) .
\end{align}

Using the identity
\[
\th_{2,1}(\t,z)-\th_{2,-1}(\t,z) = -i \th_1(\t,2z) ,
\]
 and 
 \[-q^{1/2}y\,\th_1(\t,z+\t) = \th_1(\t,z)\]
we can rewrite the above expression as
\begin{align}\label{EG_UM_relation1}
{\bf EG}(\t,z;K3) =  Z^{\Nrt,S} (\t,z) + \frac{1}{2m} \sum_{a, b\in \ZZ/m\ZZ} q^{a^2} y^{2a}\;  \phi^\Nrt\big(\t,\frac{z+a\t+b}{m}\big)  
\end{align}
for $\Nrt=\, A_1^{24}$, where $\phi^\Nrt$ is the function defined in \eq{def:phi} that encodes the umbral moonshine mock modular form $H^\Nrt$.
In the above, for a root system $\Nrt$ that is the union of simply-laced root systems with the same Coxeter number $m$ (cf. \eq{def:Em})
$$\Nrt =  A_{m-1 }^{d_A}  D_{m/2+1}^{d_D} (E^{(m)})^{d_E},   $$  
we write 
\be
 Z^{\Nrt,S}  = {d_A}  Z^{A_{m-1}} + d_D Z^{D_{m/2+1}}  + d_E Z^{E^{(m)}} ,
\ee 
corresponding to a collection of non-interacting ADE theories with the total Hilbert space given by the direct sum of the Hilbert spaces of the component theories.

In other words, instead of interpreting the two contributions to the $K3$ elliptic genus as that of the BPS and that of the non-BPS ${\cal N}=4$ multiplets, one might interpret them as the contribution from the 24 copies of $A_1$-type surface singularities and the ``umbral moonshine" contribution given by the umbral moonshine mock modular forms $H^\Nrt$ with $\Nrt=A_1^{24}$. 

The first surprise we encounter is that such an interpretation actually holds for all  23 cases of umbral moonshine.
In particular, the equality \eq{EG_UM_relation1} is valid not only for the case $\Nrt=A_1^{24}$ but also for all other 22 cases corresponding to all the 23 Niemeier lattices $L^\Nrt$.
The detailed proof will be supplied in Appendix \ref{app:Proof}. 
Put differently, corresponding to the 23 Niemeier lattices $L^\Nrt$ we have 23 different ways of separating ${\bf EG}(K3)$ into two parts.
On the one hand, by replacing the Niemeier root system $\Nrt$ with the corresponding configuration of singularities, we obtain a contribution to the $K3$ elliptic genus by the singularities.
On the other hand, the umbral moonshine construction attaches a  mock Jacobi form $\f^\Nrt$ to every $L^\Nrt$, which gives the rest of ${\bf EG}(K3)$ after a summation procedure reminiscent of the ``orbifoldisation" formula for the elliptic genus of orbifold SCFTs \cite{kawai_elliptic_1993}. 

Recall that in umbral moonshine for a given Niemeier lattice $L^\Nrt$, the mock Jacobi form $\f^\Nrt$ is conjectured to encode the graded dimension of an infinite-dimensional module $K^\Nrt$ of the umbral finite group $G^\Nrt$. The existence of such a module is supported by the construction of the other mock Jacobi forms $\f^\Nrt_g$ for the other (non-identity) conjugacy classes $[g]$ of the umbral group $G^\Nrt$ (cf. \eq{def:umbral group}), that are conjectured to encode the graded characters of $K^\Nrt$. Given the above relation between the $K3$ elliptic genus and the mock modular form  $H^\Nrt=H^\Nrt_g$ for $[g]$ being the identity class, a natural question is whether a $K3$ interpretation also exists for other mock modular forms  $H^\Nrt_g$ corresponding to other conjugacy classes of the group $G^\Nrt$. 

To discuss the relation between the graded characters in umbral moonshine and the elliptic genus of $K3$, let us first discuss how the equality \eq{EG_UM_relation1} might be ``twined" in the presence of a non-trivial group element. 
On the left-hand side (the $K3$ side) of the equation is the elliptic genus, defined in terms of the Ramond-Ramond Hilbert space ${\cal H}_{{\cal T},RR}$ of the underlying supersymmetric sigma model ${\cal T}$ as in \eq{def:elliptic_genus}. 
In the event that every Hilbert subspace ${\cal H}_{h,j;{\cal T},RR} \subset {\cal H}_{{\cal T},RR}$, consisting of states with the same $L_0, J_0$ eigenvalues $h$ and $j$,  is a representation of the cyclic group generated by $g$, or that $g$ acts on the theory and commutes with the superconformal algebra in other words, we can define the so-called ``twisted elliptic genus" as the graded character 
\be\label{def:twined_elliptic_genus}
{\bf EG}_g(\t,z;K3) = \tr_{{\mathcal H}_{{\cal T},\text {RR}}}\Big(g \,(-1)^{F_R+F_L} y^{J_0} q^{H_L}\bar q^{H_R}  \Big). 
\ee

Let us now turn to the right-hand side (the umbral moonshine side) of the equation. 
Assuming the (linear) relevance of the umbral moonshine module $K^\Nrt$ for the calculation of  ${\bf EG}(K3)$, the unique way to twine the second term
\[
 \sum_{a, b\in \ZZ/m\ZZ} q^{a^2} y^{2a}\;  \phi^\Nrt\big(\t,\frac{z+a\t+b}{m}\big) 
\]
is to replace it with 
\[
 \sum_{a, b\in \ZZ/m\ZZ} q^{a^2} y^{2a}\;  \phi_g^\Nrt\big(\t,\frac{z+a\t+b}{m}\big) 
\]
where $ \phi_g^\Nrt$ is defined in \eq{def:phi_g}. This is equivalent to replacing the graded dimension of the module $K^\Nrt$ with its graded character. 
What remains to be twined is the first term in \eq{EG_UM_relation1}, the contribution from the configuration of singularities stipulated by the root system $\Nrt$ of the Niemeier lattice $L^\Nrt$. For an element $g$ of the umbral group $G^\Nrt$, consider its action on the rank 24 root system $X$.
In the case that $g$ simply permutes the different irreducible components of its root system, it is easy to write down the twining of the singularity part $Z^{\Nrt,S}$ of ${\bf EG}(K3)$: $Z_g^{\Nrt,S}$ is simply given by the contribution from the irreducible components of $\Nrt$ that are left invariant by the action of $g$. For instance, for $\Nrt=A_1^{24}$, consider the order 2 element $g$ of the umbral group $G^\Nrt = M_{24}$ whose action on $L^\Nrt$ is to exchange 8 pairs of $A_1$ root systems and leave the other 8 copies of $A_1$ invariant when restricted to the root vectors of $L^\Nrt$. In this case the twined singularity part of the elliptic genus is simply $Z^{\Nrt,S}_g = 8 Z^{A_1,S}$. It can also happen that $g$ also involves a non-trivial automorphism of the individual irreducible components of the root system, such as the  $\ZZ/2\ZZ$ symmetry of the $A_n$, $n>1$ Dynkin diagram and the $\ZZ/3\ZZ$ symmetry of the $D_4$ Dynkin diagram. In this case the computation for $Z^{\Nrt,S}_g$ is more involved and will be discussed in Appendix \ref{sec:compute_twining}. 
Combining the two parts, we can now define the twining for the right-hand side (the umbral moonshine side) \eq{EG_UM_relation1} which we denote by 
\be\label{def:Zg_UM}
Z^{\Nrt}_g(\t,z) = Z^{\Nrt,S}_g(\t,z) + \frac{1}{2m} \sum_{a, b\in \ZZ/m\ZZ} q^{a^2} y^{2a}\;  \phi_g^\Nrt\big(\t,\frac{z+a\t+b}{m}\big). 
\ee

The second surprise is that these twining functions given by umbral moonshine precisely reproduce the  elliptic genus twined by a geometric symmetry of the underlying $K3$ surface whenever the latter interpretation is available, a fact we will now explain. 
The symmetries of a $K3$ surface $M$ that are of interest for the purpose of studying the elliptic genus are the so-called finite symplectic automorphisms of $M$, as we need to require the symmetry to preserve the hyper-K\"ahler structure in order for it to commute with the ${\cal N}=4$ superconformal algebra. 
As we will discuss in \S\ref{Geometric Interpretation}, a necessary condition for a subgroup $G\subseteq G^\Nrt$ to admit such an interpretation as the group of finite symplectic automorphisms of a certain $K3$ surface is that it has at least 5 orbits and 1 fixed point on the 24-dimensional representation of $G^\Nrt$. 
See \cite{Kondo} for a proof by S. Kond{\=o} utilising the previous results by V. Nikulin \cite{Nikulin1980,MR0409904}, and \cite{Nikulin:2011} for a more refined  analysis.

For convenience, above and in the rest of the paper we will simply refer to the 24-dimensional representation that encodes the action of $G^\Nrt$ on $\Nrt$ 
as ``the 24-dimensional representation" of $G^\Nrt$. As above and in \S\ref{Geometric Interpretation}, this representation is also the relevant one when describing the action of various subgroups of $G^\Nrt$ on the $K3$ cohomology lattice, via the embedding of its sub-lattice into $L^\Nrt$.
The action of an element $g\in G^\Nrt$ on the  24-dimensional representation is encoded in the 24 eigenvalues, or equivalently its ``24-dimensional cycle shape" 
\be\label{def:cycle_shape}
\Pi_g^\Nrt=\prod_{i}^k \ell_i^{m_i}, \;\text{where  } {m_i}\in \ZZ_{>0} ,\; 0<\ll_1<\dots<\ll_k \; \text{and } \sum_{i}^k m_i\ll_i=24,
\ee
where the relation between the cycle shape and the eigenvalues $\l_1,\dots,\l_{24}$ is given by 
\be
\prod_{i}^k (x^{\ell_i}-1)^{m_i} = (x-\l_1) \cdots  (x-\l_{24}). 
\ee
We will say that an element $g\in G^{\Nrt}$ satisfies  the ``geometric condition" if it satisfies the criterium of Mukai, namely when it has at least 5 orbits ($\sum_{i}^k m_i \geq 5$) and one fixed point ($\ll_1=1$) on the 24-dimensional representation.

Moreover, this implies that $G$ must be (isomorphic to) a subgroup of one of the 11 maximal subgroups of $M_{23}$ listed in \cite{Mukai} and provides an alternative proof of Mukai's theorem \cite{Mukai}. Conversely, given any $G^\Nrt$ among the 23 umbral groups  and for any element $g\in G^\Nrt$ satisfying the geometric condition, there exists a $K3$ surface $M$ whose finite group of symplectic automorphisms has a subgroup isomorphic to $\langle g \rangle$. 
This can be shown using the global Torelli theorem and in fact holds not just for the Abelian groups but also for all 11 maximal subgroups of $M_{23}$. See the Appendix by S. Mukai in \cite{Kondo}.  

As a result, for any of the 23 $G^\Nrt$ for any element $g\in G^\Nrt$ satisfying the geometric condition, one can compute  ${\bf EG}_g(K3)$ geometrically by 
considering the supersymmetric sigma model on a $K3$ surface with $\langle g \rangle$ symmetry. 
Note that the latter is well-defined because of the uniqueness of the $\langle g \rangle$ action.  
To be more precise, it was shown in \cite{MR544937} that if $G_i \cong \langle g \rangle$ acts on a $K3$ surface $M_i$ faithfully and symplectically ($i=1,2$), 
then there exists a lattice isomorphism $\a : H^2(M_1,\mathbb Z) \to H^2(M_2,\mathbb Z)$ preserving the intersection forms such that $\a\cdot G_1 \cdot \a^{-1} = G_2$ in $H^2(M_2,\mathbb Z)$ (see \cite{MR2926486} for a generalisation of this result to many non-Abelian groups).
Together with the global Torelli theorem, which states that any lattice isomorphism $\varphi^\ast: H^2(M,\mathbb Z) \to H^2(M',\mathbb Z)$ between the second cohomology groups of two $K3$ surfaces that preserves the Hodge structure and the effectiveness of the cycles is induced by a unique isomorphism $\varphi: M \to M' $, this shows the uniqueness of the symplectic action of $\langle g \rangle$ on $K3$ and thereby that of ${\bf EG}_g(K3)$.

On the other hand, using the prescription of  umbral moonshine \eq{def:Zg_UM} one can compute $Z^{\Nrt}_g$. 
The first non-trivial fact is that, whenever $g_1 \in G^{\Nrt_1}$ and $g_2 \in G^{\Nrt_2}$ both satisfy the geometric condition and moreover 
have the same 24-dimensional cycle shape $\Pi^{\Nrt_1}_{g_1}=\Pi^{\Nrt_2}_{g_2}$, we obtain
\be
Z^{\Nrt_1}_{g_1}=Z^{X_2}_{g_2}
\ee
despite the fact that they are defined in a very different way and each consists of two very different contributions (cf. \eq{def:Zg_UM}). 
Second, the result also coincides with the geometrically twined  elliptic genus for a $K3$ admitting $\langle g \rangle$-symmetry 
\be
Z^{\Nrt}_{g}  = {\bf EG}_g(K3)
\ee
whose induced action on 24-dimensional representation is isomorphic to that of $g\in G^\Nrt$.

For the conjugacy classes $g \in G^\Nrt$ that do not satisfy the geometric condition, the interpretation of the function $Z^\Nrt_g$ is much less clear, similar to the situation in the $M_{24}$-moonshine. 
Just like the more familiar case when $\Nrt =A_1^{24}$ \cite{Gaberdiel2011}, some of them correspond to SCA-preserving symmetries of certain SCFT ${\cal T}$ in the same moduli space as that of $K3$ sigma model, while some of them don't. 
We will discuss their interpretation in \S\ref{Conclusion and Discussion}. 
The explicit formulas for the $Z^\Nrt_g$ for all the conjugacy classes $[g]\subset G^\Nrt$ for all 23 $\Nrt$ can be found in Appendix \ref{Data}.

\section{Geometric Interpretation}
\label{Geometric Interpretation}

The result of the previous section suggests that it can be fruitful to study the symmetries of (the non-linear sigma models on) different $K3$ surfaces with different configurations of rational curves in a different framework corresponding to the 23 different cases of umbral moonshine. 
In this section we will see how, on the geometric side, this has  in fact been implemented in a recent analysis of the relation between the $K3$ Picard lattice, $K3$ symplectic automorphisms, and the Niemeier lattices \cite{Nikulin:2011,Nikulin:2014}. 
On the one hand, this provides a geometric interpretation of the results in this paper. 
On the other hand, one can view our results as a moonshine manifestation and extension of the geometric analysis in \cite{Nikulin:2011}. 

To discuss this interpretation, let us first briefly review the result in \cite{Nikulin:2011}, in which 
Nikulin advocates a more refined study of the geometric and arithmetic properties of $K3$ surfaces by introducing an additional marking using Niemeier lattices.
(Note that the idea of using the specific Niemeier lattice that has root lattice $A_1^{24}$ to mark the K3 lattice was first proposed and concretely realised in \cite{Taormina:2011rr}.)
Usually, to specify a ``{marking}" of a $K3$ surface $M$ is to specify an isomorphism between the rank 22 lattice $H^2(M,\ZZ)$ and the unique (up to isomorphism) even unimodular lattice $\G_{3,19}\cong 2E_8(-1)\oplus 3U$ of signature (3,19), where $U$ is the hyperbolic lattice $U=\big(\begin{smallmatrix}  0&1\\ 1& 0 \end{smallmatrix}\big)$ \footnote{The  ``$(-1)$" means that we multiply the lattice bilinear form by a factor of $-1$. This $(-1)$ comes from the fact that the signature of the $K3$ cohomology lattice is mostly negative while the usual convention for the signature of the simply-laced root system and hence the Niemeier lattices is positive definite. The same goes for the $(-1)$ factor in the definition of $S_M$ below.}. 
To introduce an additional marking by Niemeier lattices, on top of the marking described above, an important ingredient is the Picard lattice 
\[
{\rm Pic}(M)  = H^2(M,\ZZ) \cap H^{1,1}(M)
\]
of $M$. The real space $H^{1,1}(M,\RR)$ has signature $(1,19)$ and the Picard lattice is either: {{a}.} negative definite with $0 \leq {\rm rk}\,({\rm Pic}(M))  \leq 19$ ; {b.} hyperbolic of signature $(1, {\rm rk}\,({\rm Pic}(M)) -1)$ and with $1 \leq  {\rm rk}\,({\rm Pic}(M))  \leq 20$; c. semi-negative definite with a null direction and with $1 \leq {\rm rk}\,({\rm Pic}(M))  \leq 19$. The condition {b.} holds if and only if $M$ is algebraic. On the other hand, a generic non-algebraic $K3$ suface satisfies the first condition. Unless differently stated, we will focus on these two, the ``generic" {(a.)} and the ``algebraic" {(b.)}, cases. 

To obtain an additional marking of $M$ by a Niemeier lattice, consider the maximal negative definite sublattice of the Picard lattice, denoted by $S_M(-1) \subseteq {\rm Pic}(M)$. To be more explicit, in the generic case we have simply $S_M(-1) = {\rm Pic}(M)$, while in the algebraic case $S_M(-1) =h^\perp_{{\rm Pic}(M)}$ is the orthogonal complement in the Picard lattice of the one-dimensional sublattice generated by the primitive $h\in {\rm Pic}(M)$ with $h^2>0$ corresponding to a {nef} divisor on $M$.
Using the properties of the Torelli period map, one can show that a lattice $S_M$ may arise in the above way from a $K3$ surface $M$ if and only if $S_M(-1)$ admits a primitive embedding into $\G_{3,19}$, a condition that can be further translated into more concrete terms using the lattice embedding results in \cite{Nikulin1980}.

We say $\iota_{M,\Nrt}$ is a marking of the $K3$ surface $M$ by the Niemeier lattice $L^\Nrt$ if 
$\iota_{M,\Nrt}: S_M \to L^\Nrt$
is a primitive embedding of  $S_M$ into $L^\Nrt$. The first result of  \cite{Nikulin:2011} states that every $K3$ surface admits a marking by (at least) one of the 23 Niemeier lattices. 
This can be shown using the fact that $S_M(-1)$ admits a primitive embedding into $\G_{3,19}$ and the embedding theorem in \cite{Nikulin1980}\footnote{The trick of considering  $S_M\oplus A_1$, also used in \cite{Kondo} to prove Mukai's theorem, is  employed here to exclude the Leech lattice.}. 
We will denote by $\til S_M$ the image of $ S_M$, and $(\til S_M)^\perp_{L^\Nrt}$ by its orthonormal complement in $L^\Nrt$. 

The second result, demonstrating the importance of {\it all} 23 Niemeier lattices for the study $K3$ surfaces, proves that for every $L^\Nrt$ with the exception of $\Nrt=A_{24}$ and $\Nrt=A_{12}^2$, there exists a $K3$ surface that can only be marked using $L^\Nrt$ and not by any other Niemeier lattice. It was also conjectured in \cite{Nikulin:2011} that the same statement also holds for $\Nrt=A_{24}$ and $\Nrt=A_{12}^2$. 
In particular, from this point of view the case $\Nrt=A_{1}^{24}$ is not more special than any other of the 22 cases.
 The third result on the additional Niemeier marking states that, for any $L^\Nrt$,  any primitive sublattice of $L^\Nrt$ which can be primitively embedded into $\G_{3,19}(-1)$ arises from the Picard lattice ${\text{Pic}}(M)$ in the way described above for a certain $K3$ surface $M$. 

The above three results show that the additional marking of $K3$ lattices is general and universally applicable. 
Now we will see that such an extra marking is also useful. In \cite{Nikulin:2011}, two applications of the Niemeier marking are discussed. 
As we will see, both are crucial for the geometric interpretation of our results.
The first application is to use the Niemeier marking to constrain the configuration of smooth rational curves in a $K3$ surface: 
for the generic cases, a $K3$ surface $M$ that can be marked by $L^\Nrt$ has the configuration of all smooth rational curves given by $\Nrt\cap S_M$; for the algebraic cases, this holds modulo multiples of the primitive nef element. In particular, if one thinks of the rational curves as arising from the minimal resolutions of the du Val singularities, then the singularities have to be given by a sub-diagram of the Dynkin diagram corresponding to $\Nrt$. 
The second application involves studying the symmetries of $K3$. 
If $M$ is a $K3$ surface of the generic or the algebraic type and $M$ admits a marking by $L^\Nrt$, then the finite symplectic automorphism group $G_M$ of $M$ is a subgroup of $G^\Nrt$. More precisely, we have $$G_M= \{g\in G^\Nrt\lvert g v = v \;\;{\text{for all }} v \in (\til S_M)_{L^\Nrt}^\perp\}.$$
 In the other direction, $G\subset G^\Nrt$ is the finite symplectic automorphism group of some $K3$ surface if the orthonormal complement $(L^\Nrt)_G\subset L^\Nrt$ of the fixed point lattice $\{v\in  L^\Nrt\lvert g v= v \;\;{\text{for all }} g\in G\}$ can be primitively embedded into $\G_{3,19}(-1)$. Such $G\subset G^\Nrt$ that  arise from $K3$ symmetries have been computed in \cite{Nikulin:2011} for all 23 $L^\Nrt$. In particular, it is easy to see that they indeed satisfy the geometric condition mentioned in \S\ref{Singularities K3 Elliptic Genus}: they must have at least 5 orbits on the 24-dimensional representation and at least 1 fixed point. 

From the above two applications, we see that the marking by Niemeier lattices facilitates a more refined study of $K3$ geometry by labelling  a $K3$  surface by one of the Niemeier lattices $L^\Nrt$ via marking. This labelling is, as explained above, sometimes unique and sometimes not. It tends to be  unique when the $K3$ surface has very large symmetry --  the type of $K3$ surfaces especially of interest to us. 
In the above two applications, the two most important pieces of data associated to the Niemeier lattice $L^\Nrt$ for the construction of umbral moonshine -- the root system $X$ and the umbral group $G^\Nrt$ -- acquire the meaning of the ``enveloping smooth rational curve configuration" and the ``enveloping symmetry group" respectively, for all the $K3$ surfaces that can be labelled by $L^\Nrt$. Employing this obvious interpretation for $\Nrt$ and $G^\Nrt$,  the contribution from the ADE singularities to the (twined) $K3$ elliptic genus  (cf. \eq{EG_UM_relation1} and \eq{def:Zg_UM}) acquires the interpretation of the contribution from the ``enveloping smooth rational curve configuration" of the (class of) $K3$ surface, while the twining given by umbral moonshine is to be interpreted as encoding the action of the ``enveloping symmetry group" on the non-linear sigma model. 

Before closing the section, let us give a few examples to illustrate the above discussion. 
Consider a $K3$ surface $M$ with 16 smooth rational curves giving the root system $A_1^{16}$, generating a primitive sublattice $\Pi_{K}$ of ${\text{Pic}}(M)$. 
It is known that such a $K3$ surface is a Kummer surface, {\it i.e.} a resolution of $T^4/\ZZ_2$ by replacing the 16 $A_1$ du Val singularities with 16 rational curves \cite{Nik_Kummer}. Note that the $K3$ is not necessarily algebraic since the $T^4$ can be non-algebraic.
From the above discussion we see that $M$ can only be marked by the Niemeier lattice $L^\Nrt$ with $\Nrt = A_1^{24}$ and hence its finite symplectic automorphism group is a subgroup of $M_{24}$. More precisely, it is a subgroup of $\{g\in M_{24}| g(\Pi_{K}) =\Pi_{K}  \}$.
Similarly, let's consider as the second example a $K3$ surface $M$ with 18 smooth rational curves giving the root system $A_2^{9}$. 
It can arise in the Kummer-type construction, where we consider the minimal resolution of the nine $A_2$ type singularities of $T^4/\ZZ_3$ (for a certain type of $T^4$ and a certain $\ZZ_3$). Similarly,  $M$ can only marked by the Niemeier lattice $L^\Nrt$ with $\Nrt = A_2^{12}$ and hence its finite symplectic automorphism group is a subgroup of $G^\Nrt\cong 2.M_{12}$. For a certain $T^4/\ZZ_6$ model, by resolving the singularities of type $A_5\oplus A_2^4 \oplus A_1^5$ we obtain a $K3$ surface that can be marked by $L^\Nrt$ with $\Nrt = A_7^2 D_5^2$. See \cite{Fujiki,Wendland:2000ry} for the detailed description of these $K3$ at the orbifold limit. From the above analysis the symmetry of this $K3$ lies in $G^\Nrt\cong Dih_4$.

\section{Discussion} 
\label{Conclusion and Discussion}

In this paper we established a relation between umbral moonshine and the $K3$  elliptic genus, thereby taking a first step in placing umbral moonshine into a geometric and physical context. 
However, many questions remain unanswered and much work still needs to be done before one can solve the mystery of umbral moonshine.
In this section we discuss some of the open questions and future directions. 

\noindent
\begin{itemize}
\item{In \S\ref{Geometric Interpretation} we have provided an interpretation of the umbral group $G^\Nrt$ as the ``enveloping symmetry group"  of the (sigma model of) $K3$ surfaces  that can be marked by the given Niemeier lattice $L^\Nrt$. It would be interesting to investigate to what extent this general idea of ``enveloping symmetry group" can be made precise and can be confirmed by combining geometric symmetries at different points in the moduli space, similar to the idea explored in \cite{Taormina:2013jza}. Abstractly, it seems rather clear that varying the moduli induces a varying primitive embedding of $S_M$ into $L^\Nrt$ and  can generate a subgroup of $G^\Nrt$ that doesn't necessarily admit an interpretation as a group of geometric symmetries of any specific $K3$ surface. As a concrete example, one family of $K3$ surfaces that that might be amenable to an explicit analysis is the torus orbifold $T^4/\ZZ_3$, where one can easily vary the moduli of the $T^4$. As discussed in  \S\ref{Geometric Interpretation}, the umbral group relevant for this family is $G^\Nrt \cong 2.M_{12}$ with $X=A_2^{12}$, analogous to the $M_{24}$ case for the torus orbifold $T^4/\ZZ_2$ studied in \cite{Taormina:2013jza}.
}
\item{Another obvious possible interpretation for the conjugacy classes $[g]$ that do not admit a geometric interpretation in the present context is as  stringy symmetries of certain $K3$ sigma models preserving the ${\cal N}=(4,4)$ superconformal symmetries that have no counterpart in classical geometry. Note that they must have at least 4 orbits in the 24-dimensional representation in order for this interpretation to be possible \cite{Aspinwall:1994rg,Gaberdiel2011}. As a result, it is clear that not all conjugacy classes of all of the 23 $G^\Nrt$ admit such a possible interpretation. 
  When a conjugacy class $[g]$ does have at least 4 orbits, often the resulting umbral moonshine twining $Z^\Nrt_g$ is observed to coincide with a known  elliptic genus ${\bf EG}_{g'}(K3)$ twined by a certain symmetry $g'$ of the non-linear sigma model whose induced action on the 24-dimensional representation is isomorphic to that of $g$, {\it i.e.} they have the same cycle shape. 
However, we have not been able to match all $Z^\Nrt_g$ with some known CFT twining results for all $[g]\subset G^\Nrt$ with at least 4 orbits. 
Moreover, for non-geometric classes $g$ the twining $Z^\Nrt_g$ is not uniquely determined by the cycle shape $\Pi^\Nrt_g$ and it can occur that  $Z^\Nrt_g\neq Z^{\Nrt'}_{g'}$ even when $\Pi^\Nrt_g=\Pi^{\Nrt'}_{g'}$. See the following point for a closely-related discussion.

Curiously, various twining functions $Z^\Nrt_g$ coincide with those obtained in the work of \cite{John_Sander}. It will be interesting to understand better the relation of the two analysis.}
\item{
It seems possible and natural to generalise the analysis in \S\ref{Geometric Interpretation} beyond the realm of geometric symmetries to include the CFT symmetries. 
To do so, one should consider the ``quantum Picard lattice" ${\text{Pic}}(M)\oplus U$ instead of ${\text{Pic}}(M)$ and consider its embedding into  $\G_{4,20}=\G_{3,19}\oplus U$ instead of  $\G_{3,19}$. The relevant symmetry groups are again subgroups of $G^\Nrt$, now with at least 4 orbits on the 24-dimensional representation. 
The analysis should amount to a combination of that in \cite{Nikulin:2011} and in \cite{Gaberdiel2011}.
However, a lack of a Torelli type theorem means some of the very strong results in \cite{Nikulin:2011} will not necessarily hold for the CFT generalisation. 
Finally, given a fixed Niemeier marking one may also generalise the ``symmetry surfing" analysis (see above) into the realm of CFT symmetries. }
\item{It would be illuminating to provide the  CFT underpinning of the separation of ${\bf EG}(K3)$ into the contribution from the singularities and the rest  \eq{EG_UM_relation1}, by for instance analysing  the twisted and untwisted fields  in the orbifold $K3$ models. 
}
\item{It would be interesting to extend the geometrical definition of elliptic genus \eq{def_eg} to non-compact spaces and obtain a geometric derivation of the CFT result \eq{def:EG_ADE}. Similarly, one should compute the geometrical twined (or equivariant) elliptic genera and compare them with the conjecture in Appendix \ref{sec:compute_twining}. }
\item{The map \eq{def:Zg_UM} from the umbral moonshine function $H^\Nrt_g$ (or equivalently $\f^\Nrt_g$) to the weak Jacobi form $Z^\Nrt_g$ is a projection: the summing over the torsion points projects out terms that would have corresponded to states with fractional $U(1)$ charges. 
In particular, determining a $G^\Nrt$-module for the set of weak Jacobi forms $Z^\Nrt_g$ is in general not sufficient to construct the $G^\Nrt$-module $K^\Nrt$ underlying  $H^\Nrt_g$. 
It is hence important to gain a better understanding about the physical origin of this projection. Its form is very reminiscent of  the Landau--Ginzburg description of the non-linear sigma model and we are currently investigating the relation between umbral moonshine and Landau--Ginzburg type theories. 
}
\item{The above fact suggests that the full content of umbral moonshine might go well beyond the realm of $K3$ sigma models, and to explain the origin of umbral moonshine we might need to go beyond CFT. It has been suggested that  Mathieu moonshine has imprints in a variety of string theory setups (see for instance \cite{Cheng2010_1,Harvey:2013mda,Cheng:2013kpa,Persson:2013xpa,Harrison:2013bya,Harvey:2014cva}). Analogously, for all 23 cases of umbral moonshine, it would be interesting to explore the possible string theoretic extension of the current result.  }
\end{itemize}

\section*{Acknowledgements}

We would like to thank John Duncan, Sameer Murthy, Slava Nikulin, Anne Taormina, Jan Troost, Cumrun Vafa, Dan Whalen and in particular Shamit Kachru, for helpful discussions. 
MC would like to thank Stanford University and Cambridge University for hospitality.
SH is supported by an ARCS Fellowship. We thank the Simons Center for Geometry and Physics for hosting the programme ``Mock Modular Forms, Moonshine, and String Theory", where this project was initiated.

\appendix

\section{Special Functions}\label{sec:modforms}

First, we define the {\em Jacboi theta functions} $\th_i(\t,z)$ as follows.
\begin{align}\label{def:JacTheta}	\th_1(\t,z)
	&= -i q^{1/8} y^{1/2} \prod_{n=1}^\infty (1-q^n) (1-y q^n) (1-y^{-1} q^{n-1})\\\notag
	\th_2(\t,z)
	&=  q^{1/8} y^{1/2} \prod_{n=1}^\infty (1-q^n) (1+y q^n) (1+y^{-1} q^{n-1})\\\notag
	\th_3(\t,z)
	&=  \prod_{n=1}^\infty (1-q^n) (1+y \,q^{n-1/2}) (1+y^{-1} q^{n-1/2})\\\notag
	\th_4(\t,z) 
	&=  \prod_{n=1}^\infty (1-q^n) (1-y \,q^{n-1/2}) (1-y^{-1} q^{n-1/2})
\end{align}

In particular we will use the transformation of $\th_1$ under the Jacobi group 
\begin{align} \notag
\th_1(\t,z) & =-\th_1(\t,-z)  \\ \notag
& = e(-\tfrac{1}{2}\tfrac{z^2}{\t}) (i\t)^{-1/2} \th_1(-\tfrac{1}{\t},\tfrac{z}{\t}) \\\notag
&= e(-1/8) \,\th_1(\t+1,z) \\\label{trans_theta1} 
& = (-1)^{\l+\m} e(\tfrac{1}{2}(\l^2 \t+ 2\l z ))  \th_1(\t,z+\l\t+\m) .
\end{align}

Second, we introduce the theta functions 
 \be\label{def:theta}
 \th_{m,r}(\t,z) = \sum_{ k = r\! \pmod{2m}}\, q^{k^2/4m} y^k.
 \ee
 for $m \in \ZZ_{>0}$
 which satisfy 
 \[
  \th_{m,r}(\t,z) = \th_{m,r+2m}(\t,z) =  \th_{m,-r}(\t,-z) .
   \]
 The theta function $ \th_m = (\th_{m,r})$, $r\in \ZZ/2m\ZZ$, is a vector-valued Jacobi form of weight 1/2 and index $m$ satisfying
 \begin{align}\notag
  \th_m(\t,z) & =\sqrt{\frac{1}{2m}} \sqrt{\frac{i}{\t}} \,\ex(-\tfrac{m}{\t}z^2)\, {\cal S}_\th.\th_m(-\tfrac{1}{\t},\tfrac{z}{\t}) \\ \notag
  & = {\cal T}_\th. \th_m(\t+1,z) \\ 
  & =  \th_m(\t,z+1) = \ex(m(\t+2z)) \th_m(\t,z+\t) ,
 \end{align} 
 where the ${\cal S}_\th$ and ${\cal T}_\th$ matrices are $2m\times 2m$ matrices with entries
 \be\label{multiplier_theta}
 ({\cal S}_\th)_{r,r'} = \ex(\tfrac{rr'}{2m}) \ex(\tfrac{-r+r'}{2}) \quad,\quad ({\cal T}_\th)_{r,r'} = \ex(-\tfrac{r^2}{4m}) \,\d_{r,r'} .
 \ee

For later use we also introduce some weight two modular forms for the Hecke congruence subgroups 
\begin{gather}\label{def:hecke_congruence}
     \Gamma_0(N)
     =\left\{
  \bem
         a & b \\
         cN & d \\
       \eem\mid
     a,b,c,d\in \ZZ,\,ad-bcN=1, 
     \right\}.
\end{gather}
including $\L_N\in M_2(\G_0(N))$ for all $N\in \ZZ_{>0}$
\bea\label{Eisenstein_form}
\L_N(\t)&=&N\, q\pa_q\log\left(\frac{\eta(N\tau)}{\eta(\tau)}\right)\\\notag&=&\frac{N(N-1)}{24}\left(1+\frac{24}{N-1}\sum_{k>0}\s(k) (q^k -N q^{Nk})\right),
\eea
where $\s(k)$ is the divisor function $\s(k)=\sum_{d\lvert k}d$.
For $N=44$ we will need the unique weight two newform
\bea\notag
f_{new}^{44}&=&q+q^3 - 3q^5 + 2q^7-2q^9 - q^{11} - 4q^{13} - 3q^{15} + 6q^{17} + 
\ldots
\eea

Finally we discuss Jacobi forms following \cite{eichler_zagier}. For every pair of integers $k$ and $m$, we say a holomorphic function $\f: \HH \times \CC \to \CC$ is an {\em (unrestricted) Jacobi form} of weight $k$ and index $m$ for the Jacobi group $\SL_2(\ZZ)\ltimes \ZZ^2$ if it satisfies 
\begin{align} \label{elliptic}
\f(\t, z)&= e( m(\l^2 \t + 2\l z)) \, \f(\t, z+\l \t +\m)  \\\label{modular}
 &= e(-m \tfrac{c z^2}{c\t+d})\, (c\t+d)^{{-k}} \f(\tfrac{a\t+b}{c\t+d},\tfrac{z}{c\t+d}). 
\end{align}

The invariance of $\f(\tau,z)$ under $\tau \rightarrow \tau+1$ and $z \rightarrow z+1$ implies a Fourier expansion
\be\label{eqn:forms:jac:FouExp}
\f(\t,z) = \sum_{n,r \in \ZZ} c(n,r) q^n y^r 
\ee
for $q=e(\t)$ and $y=e(z)$,
and the elliptic transformation can be used to show that $c(n,r)$ depends only on the {\em discriminant} $D=r^2-4mn$ and on $r ~{\rm mod}~ 2m$. An unrestricted Jacobi form is called a {\em weak Jacobi form} when the Fourier coefficients satisfy $c(n,r)=0$ whenever $n< 0$.
See, for instance, \cite{Cheng:2013wca} for an introduction of Jacobi forms following \cite{eichler_zagier}.

\section{Calculations and Proofs}
\label{app:Proof}

\subsection{ Proof of \eq{EG_UM_relation1}}

The aim of this subsection is to provide more details on the elliptic genus computed in \S\ref{The Elliptic Genus of Du Val Singularities} and to prove the identity \eq{EG_UM_relation1} for all 23 Niemeier lattices $L^\Nrt$. 

As we mentioned in the main text, the Cappelli--Itzykson--Zuber matrices govern the spectrum of ${\cal N}=2$ minimal models as well as the mock modularity of  mock modular forms featuring in umbral moonshine. 
Explicitly, the matrices $\O^\Srt$ labelled by the root system $\Srt$ is given in Table \ref{ADE1}, where we have introduced for each divisor $n$ of $m$ the following matrices
\begin{align}\label{def:OmegaMatrices}
\Omega_{m}(n)_{r,r'} = \begin{cases} 1 &\text{if $r+r' = 0$ mod $2 n$ and $r-r' = 0$ mod ${2m}/{n}$,} \\ 
0 &{\rm otherwise},
\end{cases}
\end{align}
\begin{table}
\captionsetup{font=small}
\centering
\begin{tabular}{CCCC}\toprule
 \Srt 
 & \O^\Srt \\\midrule
A_{m-1} 
& \O_m{(1)} 		\vspace{0.2em}\\
D_{{m}/{2}+1} 
&  \O_m{(1)}+ \O_m{(m/2)}	\vspace{0.2em}\\
E_6 
 & \O_{12}{(1)}+ \O_{12}{(4)}+ \O_{12}{(6)} 	\vspace{0.2em}\\
E_7
& \O_{18}{(1)}+\O_{18}{(6)}+\O_{18}{(9)}   	\vspace{0.2em}\\
E_8 
& \O_{30}{(1)}+\O_{30}{(6)}+\O_{30}{(10)}+\O_{30}{(15)}
\vspace{0.1em}\\
\bottomrule
\end{tabular}
\caption{\label{ADE1}{The ADE matrices $\Omega$ of Cappelli--Itzykson--Zuber \cite{Cappelli:1987xt}}. }
\end{table}

One significance of the Cappelli--Itzykson--Zuber matrices in our context is that it captures the action of the so-called {Eichler--Zagier operator} ${\cal W}_m{(n)}$, defined for every divisor $n$ of $m$ acting on a function $f: \HH \times \CC \to \CC$ as \cite{eichler_zagier} 
\be\label{Atkin--Lehner1}
( f\lvert {\cal W}_m{(n)}) \,(\t,z) = \frac{1}{n} \sum_{a,b = 0}^{n-1} e\left(m\left(\tfrac{a^2}{n^2} \t + 2 \tfrac{a}{n}z +\tfrac{ab}{n^2}\right)\right) f\left(\t,z+\tfrac{a}{n}\t+\tfrac{b}{n}\right). 
\ee
To be more precise, acting on the theta function \eq{def:theta} it satisfies 
\be\label{EZ_on_theta}
 \th_{m} \lvert {\cal W}_m{(n)}= \O_m{(n)} \cdot\th_{m} \;.
\ee 
In order to  exploit this equality in the calculation, we define the operator ${\cal W}^\Srt$ by replacing $\O_m(n)$ with ${\cal W}_m{(n)}$ in the definition of $\O^\Srt$ (cf. Table \ref{ADE1}), with the understanding that $f|\sum_i {\cal W}_m(n_i) = \sum_i f| {\cal W}_m(n_i)$.
Similarly, we define ${\cal W}^{\Srt'} = \sum_i {\cal W}^{\Srt_i}$ for a union of the simply-laced root systems $\Srt' = \cup_i \Srt_i$ where all $\Srt_i$ have the same Coxeter number. For later convenience, analogous to \eq{def:phi} we will also define
\be\label{def:phi_P}
\f^{\Srt',P}(\t,z) =  \frac{-i \th_1(\t,mz) \th_1(\t,(m-1)z) }{\eta^3(\t)\th_1(\t,z) }(\m_{m,0}\lvert {\cal W}^{\Srt'}(\t,z))
\ee
where $m$ denotes the Coxeter number of $\Srt$ as usual.

In \cite{Cheng:2013wca} a meromorphic function 
\[
\psi^{X,P} = \m_{m,0}\lvert {\cal W}^\Nrt 
\]
was defined for every Niemeier root system $X$, where $\m_{m,0}$ is given by the Appell--Lerch sum as in \eq{def:mu_0} and ${\cal W}^\Nrt$ is defined as above. Note that $\psi^X$ as a function of $z$ has in general poles at $z\in \frac{\ZZ}{m} + \frac{\ZZ}{m} \tau$. In \cite{Cheng:2013wca}, following \cite{Dabholkar:2012nd} this meromorphic function has the interpretation as the polar part of the meromorphic Jacobi form 
\[
\psi^{X} =  \m_{m,0}\lvert {\cal W}^\Nrt  - \sum_{r\in \ZZ/2m\ZZ}H^X_r \th_{m,r} 
\]
of weight 1 and index $m$.

First, we would like to prove  
\be\label{polar_singular}
Z^{\Srt,S}(\t,z) = \frac{1}{2m}\sum_{a, b\in \ZZ/m\ZZ} q^{a^2} y^{2a}\;  \phi^{\Srt,P}\big(\t,\frac{z+a\t+b}{m}\big) .
\ee
We will start by providing more details on the expression \eq{def:EG_minimal_coset} of the minimal model elliptic genus, which is a building block of the elliptic genus of the ADE singularities \eq{def:EG_ADE}.

Fix $m$ and let $\bar m=m-2$.
The $\hat A_1$ string functions (chiral parafermion partition function times $ \eta(\t)$, see \cite{Gepner:1987qi}) are given by $c^{r}_{s}=0$ if $r=s\pmod{2}$ and otherwise
\begin{align}\notag
c^{r}_{s}(\t)  = \frac{1}{\eta^3(\t)} \sum_{\substack{-|\a|<\b\leq |\a| \\ (\a,\b)\, {\text or}\, (\tfrac{1}{2}-\a, \tfrac{1}{2}+\b) = (\tfrac{r}{2m},\tfrac{s}{2\bar m})  \,{\text{mod}}\,\mathbb Z^2 }}\text{sgn}(\a) \,q^{m\a^2-\bar m \b^2}
\end{align} 
Note that we have shifted $r$ by one compared to the convention in, for instance, \cite{Gepner:1987qi}, \cite{kawai_elliptic_1993}.
Clearly, $r \in \mathbb Z/2m \mathbb Z$ and  $s \in \mathbb Z/2\bar m \mathbb Z$, and $c^{r}_s(\t) = -c^{-r}_s(\t) = c^{r}_{-s}(\t) $.
They can also be defined through the branching relation
\begin{align}\notag
\sum_{s\in \ZZ/2\bar m\ZZ} c^{r}_{s} \th_{\bar m,s} = \frac{\th_{m,r}-\th_{m,-r}}{\th_{2,1}-\th_{2,-1}},
\end{align} 
where we have used the theta function defined in \eq{def:theta}.
Define
\begin{align}\notag
\chi^{r}_{s,\epsilon} (\t,z)&= \sum_{k\in \mathbb Z/\bar m \mathbb Z} c^{r}_{s-\epsilon+4k}(\t)\, \th_{2m\bar m, 2s+(4k-\epsilon)m}\big(\t,\frac{z}{2m}\big).
\end{align} 
We have $\epsilon \in \mathbb Z/4\mathbb Z$, from which $\epsilon = 0, 2$ correspond to the NS and $\epsilon = 1, 3$ to the Ramond sector. 
Note that now both $r$ and $s$ in $\chi^{r}_{s,\epsilon}$ take value in $\mathbb Z/2m\mathbb Z$.

Now let 
\begin{align}\notag
\til \chi^r_s (\t,z) = \chi^{r}_{s,1} (\t,z)-\chi^{r}_{s,-1} (\t,z).
\end{align}
It is easy to check that it transforms under the elliptic transformation as
\be\label{shift_I}
\til \chi^r_{s} (\t,z+a\t+b) = (-1)^{a+b} \, e(\tfrac{sb}{m}) \, e(-\tfrac{\hat c}{2} (a^2\t +2az)) \til \chi^{r}_{s-2a}(\t,z). 
\ee
They are the Ramond sector superconformal blocks relevant for the ${\cal N}=2$ minimal models with $c=3 \frac{m-2}{m}$.

Using these building blocks, the elliptic genus of the  minimal model corresponding to the simply-laced root system $\Srt$ is then given by
\begin{align}\notag
Z_{\text{minimal}}^\Srt (\t,z) = \frac{1}{2} \sum_{0 < r,r' <m} (\O^\Srt_{r,r'}-\O^\Srt_{r,-r'})\sum_{s\in \mathbb Z/2m \mathbb Z} \til \chi^r_{s}(\t,z) \til \chi^{r'}_{s}(\bar \t,0)   .
\end{align} 
Using $\O^\Srt_{r,r'}=\O^\Srt_{-r,-r'}$, $\til \chi^r_{r'}(\t,z)=-\til \chi^{-r}_{r'}(\t,z) $ and $ \til \chi^{r}_{s}( \t,0)  = \d_{r,s}-\d_{r,-s}$, we arrive at
\begin{align}\notag
Z_{\text{minimal}}^\Srt (\t,z) = \frac{1}{2} \sum_{r,r' \in \mathbb Z/2m \mathbb Z} \O^\Srt_{r,r'} \til \chi^r_{r'}(\t,z) =\frac{1}{2}{\text{Tr}( \O^\Srt \cdot  \til \chi)}   .
\end{align}

Now we define for any $n \til n=m$, $n, \til n \in \mathbb Z$ and operator acting on a function $f: \HH \times \CC \to \CC$ as
\be\notag
f\big\lvert {\til {\cal W}}_m(n) (\t,z) = \frac{1}{n} \sum_{\substack{a,b\in \mathbb Z/ m\ZZ \\ a,b=0 \,(\til n)}}
(-1)^{a+b+ab} \ex\big( \tfrac{m-2}{2m} (a^2 \t + 2a z + ab) \big) f(\t,z+a\t+b)
\ee
Using \eq{shift_I} it is easy to check that
\be\notag
\til \chi^r_{s} \big\lvert {\til {\cal W}}_m(n) = (\O_m(n) \cdot \til \chi )^r_s = \sum_{s' \in \mathbb Z/2m \mathbb Z} \d_{s-s',0 ~(2\til n)}\d_{s+s',0 ~(2 n)} \til \chi^r_{s'}  
\ee

Finally, one can verify that
\begin{align}\notag
 \sum_{\a,\b =0}^{m-1} (-1)^{\a+\b} q^{\a^2/2} y^{\a}\big( \til \chi^r_{s} \big\lvert {\til {\cal W}}_m(n)\big) (\t,z+\a\t+\b) \,\m\big(\t,\frac{z+\a\t+\b}{m}\big) \\\notag
 = \sum_{\a,\b =0}^{m-1} (-1)^{\a+\b} q^{\a^2/2} y^{\a}  \til \chi^r_{s}(\t,z+\a\t+\b) \,\big(\m\big\lvert { {\cal W}}_m(n)\big) \big(\t,\frac{z+\a\t+\b}{m}\big). 
\end{align}
Subsequently, the identity \eq{polar_singular}  follows from the above equality and 
$$Z_{\text{minimal}}^{A_{m-1}} (\t,z)  =\tfrac{1}{2}\, {\text{Tr}}\til \chi =\frac{\th_1(t,z/m)}{\th_1(t,z(m-1)/m)}.$$

Finally we are ready to prove \eq{EG_UM_relation1}, which can be re-expressed as
\be\label{eq_EG_UM2}
{\bf EG}(\t,z;K3) =  \frac{1}{2m}\sum_{a, b\in \ZZ/m\ZZ} q^{a^2} y^{2a}\;  \phi^{\Nrt,T}\big(\t,\frac{z+a\t+b}{m}\big) 
\ee
when combined with the identity \eq{polar_singular} that we just verified and when we use the definition 
$$\phi^{\Nrt,T}(\t,z) =( \phi^{\Nrt,P} + \phi^\Nrt)(\t,z)= \frac{-i \th_1(\t,mz) \th_1(\t,(m-1)z) }{\eta^3(\t)\th_1(\t,z) }\psi^\Nrt(\t,z).$$ 
From the fact that $\psi^\Nrt$ transforms as a weight 1, index $m$ Jacobi form and using the transformation \eq{trans_theta1} of the Jacobi theta function, it is straightforward to show that the RHS of \eq{eq_EG_UM2} transforms as a weight 0, index 1 Jacobi form. Moreover, the poles of $\psi^\Nrt$ at $m$-torsion points are combined with the zeros of $\th_1(\t,mz)$ and as a result $\phi^{\Nrt,T}$ is a  holomorphic function on $\HH\times \CC$ admitting a double-expansion in powers of $q$ and $y$.  In order to show that the RHS of \eq{eq_EG_UM2} is a weight 0, index 1 weak Jacobi form, we need to prove that there is no term in its Fourier expansion with $q^n, ~n<0$. This can be shown by using the explicit formulas involving $\m_{m,0}$ and $\th_1$, combining with the fact that $H^\Nrt_r=O(q^{-r^2/4m})$ and the fact that the sum over $b$ projects out all terms with fractional powers of $y$. After showing that both sides of \eq{eq_EG_UM2} are weight 0, index 1 weak Jacobi forms, using the fact that the space of such functions is one-dimensional, the equality is proven by comparing both sides at, say, $z=0$.

\subsection{ Computing $Z^\Nrt_g$ }
\label{sec:compute_twining}

In this subsection we compute the twining function  $Z^\Nrt_g$ in \eq{def:Zg_UM}. The results of the computation are recorded in Appendix \ref{Data}. In particular, we will give the details of the computation of $Z^{X,S}_g$. As a part of the computation, we also make conjectures for the elliptic genus $Z^{\Srt,S}_h$ of du Val singularities twined by certain automorphisms $\langle h \rangle$ of the corresponding Dynkin diagram $\Srt$. 

From the action of $g\in G^\Nrt$ on the Niemeier root lattice $\Nrt$, we can divide the conjugacy classes $[g]$ into the following two types. 
In the first type, there exists an element in the conjugacy class that only permutes the irreducible components of $\Nrt$. More precisely, there exists  an element $g$ in the class that descends from an element in $ \bar G^\Nrt \subseteq G^\Nrt$, where $ \bar G^\Nrt $ is a quotient of $G^\Nrt$ and is defined by 
\begin{gather}\notag
	\bar{G}^{\rs}=\Aut(L^{\rs})/\hat{W}^{\rs},
\end{gather}
where $\hat{W}^{\rs}<\Aut(L^{\rs})$ is the subgroup of lattice automorphisms that stabilize the irreducible components of $\rs$. See Table \ref{tab:mugs} for the list of $\bar G^\Nrt$. In the second type, the action of an element in $[g]$ necessarily involves certain non-trivial automorphisms of some of the irreducible components  in $\Nrt$. See \cite{Cheng:2013wca} for a more detailed discussion.

\begin{figure}[h]
\begin{center}
\includegraphics[scale=0.3]{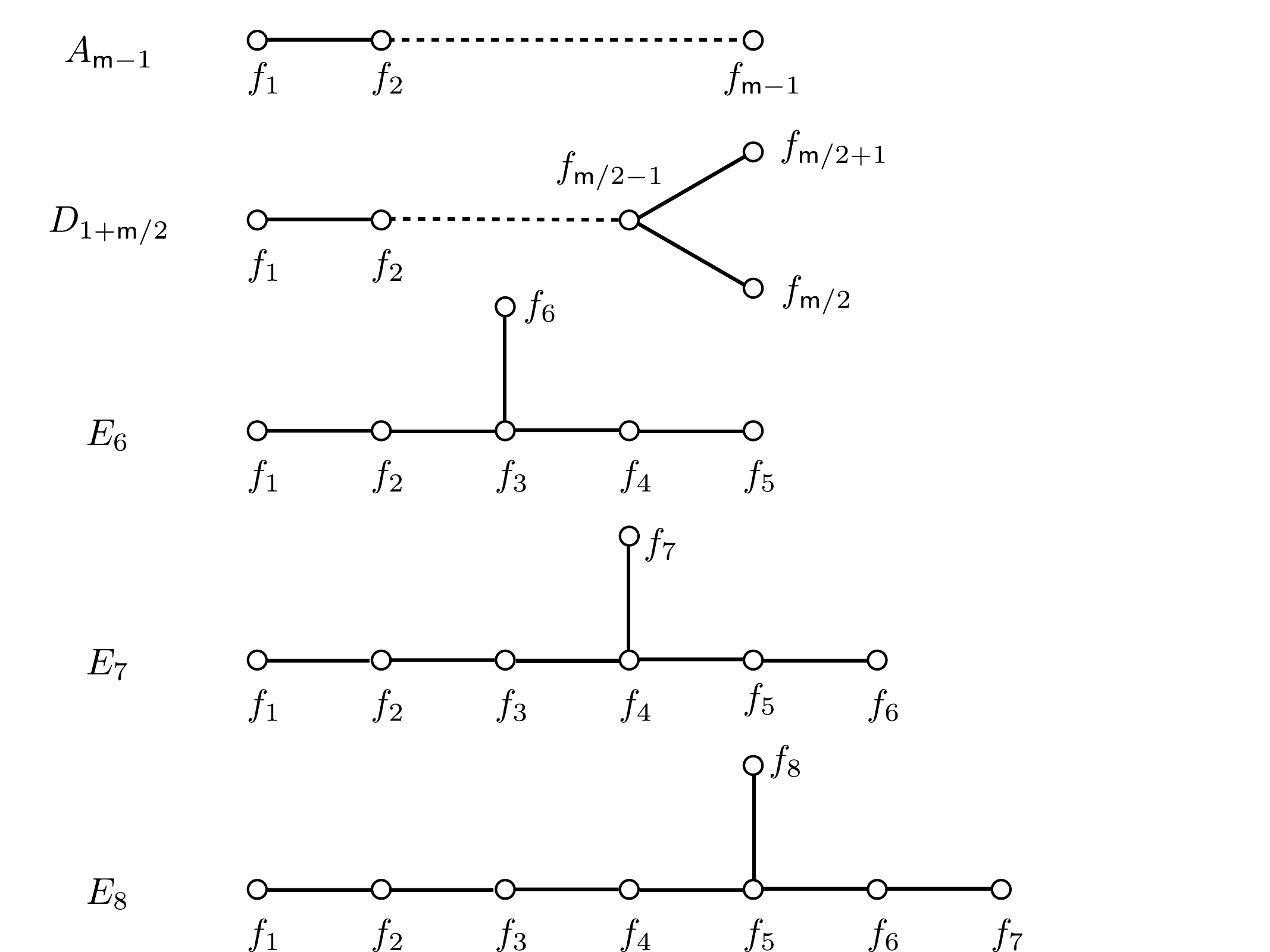}
\caption{The ADE Dynkin diagrams\label{fig:dynkin}}
\end{center}
\end{figure}

As mentioned in \S\ref{Singularities K3 Elliptic Genus}, the twined function $Z^{\Nrt,S}_g$  for a conjugacy class $[g]$ of the first type, point-wise fixing a (not necessarily non-empty) union $\Nrt_g = \cup_i \Srt_i \subset \Nrt$ of the irreducible components $\Srt_i$, is simply given by 
\[
Z^{\Nrt,S}_g = \sum_i Z^{\Srt_i,S}.
\]

In order to compute the twined function $Z^{\Nrt,S}_g$  for $[g]$ for the second type of conjugacy classes, we need to twine the elliptic genus of the (irreducible) ADE singularities by symmetries corresponding to the automorphisms of the Dynkin diagram $\Srt$. In the rest of this appendix we will propose a conjectural answer.

For the $A_{m-1}$ singularity with $m>2$ we have the $\ZZ_2$ automorphism exchanging the simple root $f_i$ with $f_{m-i}$, in the notation shown in Figure \ref{fig:dynkin}. We  conjecture that the corresponding twined elliptic genus is
\be\notag
Z^{\Srt,S}_{\ZZ_2}  = Z^{\Srt,S} \lvert {\cal W}^{(-)}\; ,\; \Srt = A_{m-1}, 
\ee
where we have defined the operator acting on a function $f:\HH\times \CC\to \CC$ as
\[
f\lvert {\cal W}^{(-)} (\t,z) = -f(\t,z+\tfrac{1}{2}).
\]

In fact, the above expression for $Z^{\Srt,S}_{\ZZ_2} $ can be deduced from the action of $\ZZ_2$ on the  eigenvectors of the appropriate Coxeter element, and similarly for the twined elliptic genus of the D- and E-type singularities discussed below.

For later use we also define the operators
\begin{align}\notag
f\lvert {\cal W}^{(3)} (\t,z) = \frac{1}{3}\sum_{a=0}^2 f(\t,z+\tfrac{a}{3})\\\notag
f\lvert {\cal W}^{(6)} (\t,z) = \frac{1}{6}\sum_{a=0}^5 f(\t,z+\tfrac{a}{6}). 
\end{align}
We remark that the above conjecture, if proven, provides a geometrical explanation of the following interesting property of the $G^\Nrt$-module $K^\Nrt$. 
It was observed and conjectured in \cite{UM,Cheng:2013wca} that in the cases where  $\Nrt$ has only A-type components ({\it i.e.} when $m-1|24$), the $G^\Nrt$-module $K^\Nrt_r$ underlying the even components of the mock modular form $H^\Nrt_{g,r}$ ($r$ even), are composed of irreducible faithful representations of $G^\Nrt$. On the other hand, the module $K^\Nrt_r$ underlying the odd components of the mock modular form $H^\Nrt_{g,r}$ ($r$ odd), are composed of $G^\Nrt$-representations that factor through $\bar G^\Nrt$. Similar considerations also apply to the cases when $\Nrt$ contains also D- and E-type components.

For the D-type  singularity different from $D_4$,  we have the $\ZZ_2$ automorphism exchanging the simple root $f_{m/2}$ with $f_{1+m/2}$, in the notation shown in Figure \ref{fig:dynkin}. We  conjecture that the corresponding twined elliptic genus is
\be\notag
Z^{\Srt,S}_{\ZZ_2} (\t,z) =  \frac{1}{m}\sum_{a, b\in \ZZ/m\ZZ} q^{a^2} y^{2a}\;  \phi_{\ZZ_2}^{\Srt,P}\big(\t,\frac{z+a\t+b}{m}\big)
\ee
for $D_{1+m/2}$ for $m\neq 6$, where
\be\label{def:phi_P}
\f^{\Srt,P}_{\ZZ_2}(\t,z) =  \frac{-i \th_1(\t,mz) \th_1(\t,(m-1)z) }{\eta^3(\t)\th_1(\t,z) }\big(-\m_{m,0}\lvert {\cal W}_m(m/2)+\m_{m,0}\lvert {\cal W}^{(-)} \big)(\t,z).
\ee

The $\Srt= D_4$ Dynkin diagram permits a $S_3$ symmetry on the roots $\{f_1,f_3,f_4\}$.  We conjecture that the corresponding twined elliptic genera are given by
\begin{align}\notag
Z^{\Srt,S}_{\ZZ_2} (\t,z) =  \frac{1}{6}\sum_{a, b\in \ZZ/6\ZZ} q^{a^2} y^{2a}\;  \phi_{\ZZ_2}^{\Srt,P}\big(\t,\frac{z+a\t+b}{6}\big)\\\notag
Z^{\Srt,S}_{\ZZ_3} (\t,z) =  \frac{1}{6}\sum_{a, b\in \ZZ/6\ZZ} q^{a^2} y^{2a}\;  \phi_{\ZZ_3}^{\Srt,P}\big(\t,\frac{z+a\t+b}{6}\big)
\end{align}
where 
\begin{align}\notag
\f^{\Srt,P}_{\ZZ_2}(\t,z) =  \frac{-i \th_1(\t,6z) \th_1(\t,5z) }{\eta^3(\t)\th_1(\t,z) }\m_{6,0}\lvert {\cal W}_{D_4,\ZZ_2}(\t,z)\\\notag
\f^{\Srt,P}_{\ZZ_3}(\t,z) =  \frac{-i \th_1(\t,6z) \th_1(\t,5z) }{\eta^3(\t)\th_1(\t,z) }\m_{6,0}\lvert {\cal W}_{D_4,\ZZ_3}(\t,z)
\end{align} 
and 
\begin{align}\notag
{\cal W}_{D_4,\ZZ_2} &= - {\cal W}_6(3) +{\cal W}^{(-)} +2{\cal W}^{(6)}  \\\notag
{\cal W}_{D_4,\ZZ_3}& = {\cal W}^{D_4} -3 {\cal W}^{(3)} .
\end{align}

The only E-type diagram with non-trivial automorphism is the $\ZZ_2$ generated by the action $f_{i}\mapsto f_{6-i}$ of $E_6$, for $1\leq i\leq 5$. We conjecture that corresponding twined elliptic genus is 
\be\notag
Z^{E_6,S}_{\ZZ_2} (\t,z) =  \frac{1}{12}\sum_{a, b\in \ZZ/12\ZZ} q^{a^2} y^{2a}\;  \phi_{\ZZ_2}^{E_6,P}\big(\t,\frac{z+a\t+b}{12}\big)
\ee
where
\be\label{def:phi_P}
\f^{E_6,P}_{\ZZ_2}(\t,z) =  \frac{-i \th_1(\t,12z) \th_1(\t,11z) }{\eta^3(\t)\th_1(\t,z) }\big(\m_{12,0}\lvert {\cal W}_{12}(6)+\m_{12,0} \lvert {\cal W}^{(-)}+\m_{12,0}\lvert
 {\cal W}_{12}(4) \lvert {\cal W}^{(-)} \big)(\t,z).
\ee

After giving the conjectural answer for the building blocks $Z^{\Srt,S}_h$ of the twining of  $Z^{\Nrt,S}_g$, we need to know how such a $g\in G^\Nrt$ acts on the Niemeier root system $\Nrt$. This is encoded in the  twisted Euler characters $\bar{\chi}^{\rs_A}$, ${\chi}^{\rs_A}$, $\bar{\chi}^{\rs_D}$, ${\chi}^{\rs_D}$ $\dots$  attached to the A-, D-, and E-components of each $\Nrt$. See \S 2.4 of \cite{Cheng:2013wca} for details and see Appendix B.2 of the same reference for the  values of such twisted Euler characters for all 23 $\Nrt$. Combining these ingredients leads to the answer for $Z^{\Nrt,S}_g$ for all conjugacy classes $[g]$ for all the umbral groups $G^\Nrt$.  This completes our computation of $Z^{\Nrt}_g$ \eq{def:Zg_UM}.

\section{The Twining Functions}
\label{Data}

In this appendix we provide the expression of $Z^{\Nrt}_g(\t,z) $  (cf. \eq{def:Zg_UM}) in terms of the function $h^\Nrt_g$: 
\be\notag
Z^{\Nrt}_g(\t,z)  =\frac{i\th_1(\t,z)^2}{\th_1(\t,2z)\eta^3(\t)} \Big\{  c^{\Nrt}_g \m_{2,0}(t,z) +h^\Nrt_g(\t) (\th_{2,-1}(\t,z) -\th_{2,1}(\t,z))\Big\}
\ee
where $c^{\Nrt}_g$ is the number of fixed point in the 24-dimensional representation of $G^\Nrt$. In other words, for the cycle shape $\Pi_g^\Nrt$ defined in \eq{def:cycle_shape}, we have $c^{\Nrt}_g = m_1$ if $\ll_1=1$ and $c^{\Nrt}_g =0$ otherwise. For instance, for $\Nrt=A_1^{24}$ and $[g]$ the identity class, the above formula gives the ${\cal N}=4$ character decomposition of ${\bf EG}(K3)$ in \eq{EG_K3_decomposition}. 

For $\Nrt=A_1^{24}$, the functions $h^\Nrt_g(\t)$ for all $[g]\subset G^\Nrt \cong M_{24}$ have been worked out in \cite{Cheng2010_1,Gaberdiel2010,Gaberdiel2010a,Eguchi2010a}. We refer to these papers, or the summary in \cite{CheDun_M24MckAutFrms,UM}. For convenience we will denote $h_g^{A_1^{24}}$ simply by $H_g$ for $[g]\subset  M_{24}$. Recall that  $H_g$ is nothing but the function discussed in \eq{def:H21} when $[g]=1A$ is the identity class of $M_{24}$. There are two cases, corresponding to $\Nrt=D_{24}$ and  $\Nrt=D_{16}E_8$, with trivial $G^\Nrt$.  As a result they are not included in the present appendix.

When $h^\Nrt_g$ coincides with $H_{g'}$ for a certain $g'$, we will simply use this identity to define $h^\Nrt_g$. When there does not exist such a $[g']\subset  M_{24}$, we write 
\be\label{h_g_explicit}
h^\Nrt_g(\t) = \frac{c^{\Nrt}_g}{24} H_{1A}(\t) - \frac{\til T^\Nrt_g(\t)}{\h(\t)^3}\;
\ee
and we will give the explicit expression for $\til T^\Nrt_g$ in the following tables  using the functions given in Appendix \ref{sec:modforms}. We also use the short hand notation $(n)^k := \eta^k(n\t)$.

\begin{table}[h]
\begin{center}
\caption{$X=A_2^{12}$}
\begin{tabular}{CCCCCLLL}
[g]&
 \Pi_g^\Nrt& h^\Nrt_g \\\midrule
1A &  1^{24}    &  H_{1A} \\ 
2A &  2^{12} & H_{2B} \\ 
4A&4^6  &H_{4C} \\
2B&  1^82^8& H_{2A}\\
2C&  {2^{12}}&H_{2B}   \\
3A& 1^63^6& H_{3A} \\
6A& 2^36^3  & \til T_{6A}^\Nrt = 3\Lambda_2+2\Lambda_3-\Lambda_4-3\Lambda_6+\Lambda_{12} \\
3B & 3^8& \til T_{3B}^\Nrt = 2 ( -4 \L_3 + \L_9 - (1)^6/(3)^2 )  \\
6B&{6^4}& \til T_{6B}^\Nrt =2{(1)^5(3)\over (2)(6)}\,\\
4B&2^44^4 &H_{4A} \\
4C&1^42^24^4&H_{4B}\\
5A&1^45^4&H_{5A}\\
10A&2^210^2  &H_{10A} & \\
12A & {12^2}  &\til T_{12A}^\Nrt=2{(1)(2)^5(3)\over (4)^2(6)}\\
6C& 1^22^23^26^2&H_{6A}\\
6D&2^36^3&h_{6D}^\Nrt = h_{6A}^\Nrt \\
8AB& 4^28^2 &\til T_{8AB}^\Nrt = 2 (2)^4(4)^2/ (8)^2 \\
8CD& 1^22^14^18^2  &H_{8A}  \\
20AB& 4^120^1&\til T_{20AB}^\Nrt =2{(2)^7(5)\over (1)(4)^2(10)}\\
11AB& 1^211^2&\til T_{11AB}^\Nrt =  ( 2\Lambda_{11}(\t) +33(1)^2(11)^2)/5 \\
22AB& 2^122^1&  \til T_{22AB}^\Nrt = (3\Lambda_2-\Lambda_4+2\Lambda_{11}-3\Lambda_{22}+\Lambda_{44})/15-{22\over 3} f_{new}^{44}+{11\over 5}(1)^2(11)^2\\
&&~~+{44\over5}(2)^2(22)^2+{88\over5}(4)^2(44)^2\\\bottomrule
\end{tabular}
\end{center}
\end{table}

\begin{table}[h]
\begin{center}
\caption{$X=A_3^8$}
\begin{tabular}{CCCCCLLL} 
[g]&  \Pi_g^\Nrt & h^\Nrt_g \\\midrule
1A&  1^{24}  &  H_{1A} \\ 
2A&  1^8{2^8} &  H_{2A} \\ 
2B& 2^{12} & H_{2B} \\ 
4A&  {2^4}{4^4}  &H_{4A} \\
4B& 4^6 & \til T_{4B}^{\Nrt} = 2\Lambda_4-3\Lambda_{8}+\Lambda_{16}- 2(1)^4(2)^2/(4)^2\\
2C& 1^82^8 &	 H_{2A} \\ 
3A& 1^63^6& 	 H_{3A} \\ 
6A& 1^22^23^26^2&H_{6A} \\ 
6BC& 2^36^3&h^{\Nrt}_{6BC} = h^{Y}_{6A}, \;Y=A_2^{12}\\ 
8A& {4^2}{8^2}&\til T_{8A}^{\Nrt} = (2\Lambda_8-3\Lambda_{16}+\Lambda_{32})/8+ 8(4)^4(16)^4/(8)^4-8 (4)^2(8)^2\\
4C&  1^42^24^4 &H_{4B} \\ 
7AB&1^37^3 &H_{7AB}\\
14AB& {1^12^17^114^1}&\til T_{14AB}^{\Nrt} =( -\L_2-\L_7+\L_{14}+28(1)(2)(7)(14))/3
 \\\bottomrule
\end{tabular}
\end{center}
\vspace{10pt}

\begin{center}
\caption{$X=A_4^6$}
\vspace{10pt}
\begin{tabular}{CCCCCLLL} 
[g]& \Pi^\Nrt_g & h^\Nrt_g \\\midrule
1A& 1^{24}    &  H_{1A} \\ 
2A&  2^{12} & H_{2B} \\ 
2B& 2^{12} & H_{2B} \\ 
2C&  {1^8}{2^8}  &H_{2A} \\
3A& 3^8 & H_{3B}\\
6A& 6^4 &	 H_{6B} \\ 
5A& 1^45^4&	 H_{5A} \\ 
10A&2^210^2& \til T_{10A}^{\Nrt} = (3\Lambda_2-\Lambda_4+2\Lambda_{5}-3\Lambda_{10}+\Lambda_{20}+40(2)^2(10)^2)/3 \\ 
4AB&  4^6&h^{X}_{4AB} = h^{Y}_{4B}, \;Y=A_3^{8}\\ 
4CD& 1^42^24^4&H_{4B}\\
12AB& 12^2 &\til T_{12AB}^{\Nrt}=2{(2)^2(6)^4(1)^2\over (3)^2(12)^2}\\
\bottomrule
\end{tabular}
\end{center}
\vspace{10pt}
\begin{center}
\caption{$X=A_5^4D_4$}\vspace{10pt}
\begin{tabular}{CCCCCCCCLLL} 
[g]&  \Pi^\Nrt_g& h^\Nrt_g \\\midrule
1A& 1^{24}    &  H_{1A} \\ 
2A&1^82^8 & H_{2A} \\ 
2B&1^82^8 & H_{2A} \\ 
4A& 1^42^24^4 & H_{4B} \\ 
3A & {1^6}{3^6}  &H_{3A} \\
6A&  1^22^23^26^2 & H_{6A}\\
8AB &1^22^14^18^2 & H_{8A} \\ 
\bottomrule
\end{tabular}
\end{center}
\end{table}

\begin{table}
\begin{center}
\caption{$X=D_4^6$}
\begin{tabular}{CCCCCCLLL} 
[g]& \Pi^\Nrt_g &h^\Nrt_g \\\midrule
1A& 1^{24}    &  H_{1A} \\ 
3A& 1^6 3^6  & H_{3A} \\ 
2A& 1^8 2^8 & H_{2A} \\ 
6A& 1^22^23^26^2  &H_{6A} \\
3B& 1^6 3^6  & H_{3A}\\
3C&  3^8  & h_{3C}^{\Nrt}=h_{3B}^{Y}, \;Y=A_2^{12}\\
4A& 2^4 4^4  & H_{4A}\\
12A& 2^14^16^112^1  & \til T_{12A}^{\Nrt} = (-2\Lambda_2+3\Lambda_4+2\Lambda_6-\Lambda_8-3\Lambda_{12}+\Lambda_{24})/4+18(2)(4)(6)(12) \\
5A&  1^45^4  & H_{5A}\\
15AB&  1^13^15^115^1  & \til T_{15AB}^{\Nrt} =(-\L_3-\L_5+\L_{15} +45(1)(3)(5)(15))/4 \\
2B&  1^8 2^8  & H_{2A} \\
2C& 2^{12}  & H_{2B}\\
4B& 1^4 2^2 4^4 & H_{4B} \\
6B& 1^2 2^2 3^2 6^2  & H_{6A} \\
6C&6^4  & h_{6C}^{\Nrt}=h_{6B}^{Y}, \;Y=A_2^{12}\\
\bottomrule
\end{tabular}
\end{center}

\vspace{10pt}
\begin{center}
\caption{$X=A_6^4$}\vspace{10pt}
\begin{tabular}{CCCCCLLL} 
[g] &  \Pi^\Nrt_g & h^\Nrt_g \\\midrule
1A & 1^{24}    &  H_{1A} \\ 
2A& 2^{12} & H_{2B} \\ 
4A& 4^{6} & H_{4C} \\ 
3AB&  {1^6}{3^6}  &H_{3A} \\
6AB&  2^36^3 & h_{6AB}^{X}=h_{6A}^{Y}=h_{6BC}^{Z}, \;Y=A_2^{12}, Z=A_3^8 \\
\bottomrule
\end{tabular}
\end{center}

\vspace{10pt}
\begin{center}
\caption{$X=A_7^2D_5^2$}\vspace{10pt}
\begin{tabular}{CCCCCCCLLL} 
[g]& \Pi^\Nrt_g & h^\Nrt_g \\\midrule
1A& 1^{24}    &  H_{1A} \\ 
2A& 1^82^8 & H_{2A} \\ 
2B& 1^82^8 & H_{2A} \\ 
2C&  1^82^8 & H_{2A} \\ 
4A & {2^4}{4^4}  &H_{4A} \\
\bottomrule
\end{tabular}
\end{center}
\end{table}

\begin{table}[h]
\begin{center}
\caption{$X=A_8^3$}
\begin{tabular}{CCCCCLLL} 
[g]&\Pi^\Nrt_g & h^\Nrt_g \\\midrule
1A&  1^{24}    &  H_{1A} \\ 
2A&  2^{12} & H_{2B} \\ 
2B& 1^82^8 & H_{2A} \\ 
2C&  2^{12}  &H_{2B} \\
3A&  3^8 & h_{3A}^{\Nrt}=h_{3B}^{Y}\;, Y=A_2^{12}\\
6A&  6^4 & \til T_{6A}^{\Nrt} = 2{(1)^5(3)\over (2)(6)}+24 (6)^4 \, \\
\bottomrule
\end{tabular}
\end{center}
\vspace{10pt}
\begin{center}
\caption{$X=A_9^2D_6$}\vspace{10pt}
\begin{tabular}{CCCCCCCLLL} 
[g]&\Pi^\Nrt_g & h^\Nrt_g \\\midrule
1A&  1^{24}    &  H_{1A} \\ 
2A& 1^82^8 & H_{2A} \\ 
4AB& 1^42^2{4^4}  &H_{4B} \\
\bottomrule
\end{tabular}
\end{center}
\vspace{10pt}

\begin{center}
\caption{$X=D_6^4$}\vspace{10pt}
\begin{tabular}{CCCCCLLL} 
[g]&\Pi^\Nrt_g &h^\Nrt_g \\\midrule
1A& 1^{24}    &  H_{1A} \\ 
2A&  2^{12} & H_{2B} \\ 
3A& 1^63^6 & H_{3A} \\
2B&1^82^8 & H_{2A} \\
4A&4^6 & H_{4C}\\
\bottomrule
\end{tabular}
\vspace{10pt}
\caption{$X=A_{11}D_7E_6$}\vspace{10pt}
\begin{tabular}{CCCCCCCCCLLL} 
[g]& \Pi^\Nrt_g &h^\Nrt_g \\\midrule
1A&  1^{24}   &  H_{1A} \\ 
2A& 1^82^8 & H_{2A} \\ 
\bottomrule
\end{tabular}
\end{center}
\end{table}

\begin{table}
\begin{center}
\caption{$X=E_{6}^4$}
\begin{tabular}{CCCCCLLL} 
[g]&\Pi^\Nrt_g &h^\Nrt_g \\\midrule
1A&  1^{24}   &  H_{1A} \\ 
2A& 1^82^8  & H_{2A} \\ 
2B& 1^82^8  & H_{2A} \\
4A&  2^44^4  & H_{4A} \\
3A&  1^63^6  & H_{3A} \\
6A&  1^22^23^26^2  & H_{6A} \\
8AB& 4^28^2  & \til T_{8AB}^{\Nrt} = (2\Lambda_8-3\Lambda_{16}+\Lambda_{32})/8+24(4)^2(8)^2+ 8(4)^4(16)^4/(8)^4 \\
\bottomrule
\end{tabular}
\end{center}
\vspace{10pt}

\begin{center}
\caption{$X=A_{12}^2$}\vspace{10pt}
\begin{tabular}{CCCCCLLL} 
[g]& \Pi_g^\Nrt &h^\Nrt_g \\\midrule
1A& 1^{24}    &  H_{1A} \\ 
2A&  2^{12} & H_{2B} \\ 
4AB&4^6 & h_{4AB}^{\Nrt}=h_{4B}^{Y}=h_{4AB}^{Z}\;,Y=A_3^8,\; Z=A_4^6  \\
\bottomrule
\end{tabular}
\end{center}
\vspace{10pt}
\begin{center}
\caption{$X=D_{8}^3$}\vspace{10pt}
\begin{tabular}{CCCCCLLL} 
[g]& \Pi^\Nrt_g &h^\Nrt_g \\\midrule
1A&  1^{24}    &  H_{1A} \\ 
2A&  1^82^8 & H_{2A} \\ 
3A& 3^8 & H_{3B} \\
\bottomrule
\end{tabular}
\end{center}
\vspace{10pt}
\begin{center}
\caption{$X=A_{15}D_9$}\vspace{10pt}
\begin{tabular}{CCCCCCCLLL} 
[g]& \Pi^\Nrt_g & h^\Nrt_g \\\midrule
1A&  1^{24}    &  H_{1A} \\ 
2A& 1^82^8 & H_{2A} \\ 
\bottomrule
\end{tabular}
\end{center}
\end{table}

\begin{table}[h]
\begin{center}
\caption{$X=A_{17}E_7$}
\begin{tabular}{CCCCCCLLL} 
[g]& \Pi^\Nrt_g &h^\Nrt_g \\\midrule
1A&  1^{24}    &  H_{1A} \\ 
2A& 1^82^8 & H_{2A} \\ 
\bottomrule
\end{tabular}
\end{center}
\vspace{10pt}
\begin{center}
\caption{$X=D_{10}E_7^2$}\vspace{10pt}
\begin{tabular}{CCCCCCLLL} 
[g]& \Pi^\Nrt_g & h^\Nrt_g \\\midrule
1A& 1^{24}    &  H_{1A} \\ 
2A &1^82^8 & H_{2A}\\ 
\bottomrule
\end{tabular}
\end{center}
\vspace{10pt}
\begin{center}
\caption{$X=D_{12}^2$}\vspace{10pt}
\begin{tabular}{CCCCCLLL} 
[g]& \Pi^\Nrt_g & h^\Nrt_g \\\midrule
1A& 1^{24}    &  H_{1A} \\ 
2A&  2^{12} & H_{2B} \\ 
\bottomrule
\end{tabular}
\end{center}
\vspace{10pt}

\begin{center}
\caption{$X=A_{24}$}\vspace{10pt}
\begin{tabular}{CCCCCLLL} 
[g]& \Pi^\Nrt_g &h^\Nrt_g \\\midrule
1A&  1^{24}    &  H_{1A} \\ 
2A&  2^{12} & H_{2B} \\ 
\bottomrule
\end{tabular}
\end{center}

\vspace{10pt}

\begin{center}
\caption{$X=E_{8}^3$}\vspace{10pt}
\begin{tabular}{CCCCLLL} 
[g]& \til \Pi_g & Z_g \\\midrule
1A&  1^{24} 
&  H_{1A} \\ 
2A& 1^{8}2^8 & H_{2A} \\ 
3A & 3^8 &  h_{3B}^{\Nrt}=h_{3B}^{Y},\;Y=A_2^{12} \\ 
\bottomrule
\end{tabular}
\end{center}
\end{table}

\clearpage 
\addcontentsline{toc}{section}{References}
\bibliographystyle{utphys}
\bibliography{UMK3}

\providecommand{\href}[2]{#2}\begingroup\raggedright\begin{thebibliography}{10}

\bibitem{conway_norton}
J.~H. Conway and S.~P. Norton, ``{Monstrous Moonshine},'' {\em Bull. London
  Math. Soc.} {\bf 11} (1979)  308~339.

\bibitem{FLMPNAS}
I.~B. Frenkel, J.~Lepowsky, and A.~Meurman, ``A natural representation of the
  {F}ischer-{G}riess {M}onster with the modular function {$J$} as character,''
  {\em Proc. Nat. Acad. Sci. U.S.A.} {\bf 81} (1984) no.~10, Phys. Sci.,
  3256--3260.

\bibitem{FLMBerk}
I.~B. Frenkel, J.~Lepowsky, and A.~Meurman, ``A moonshine module for the
  {M}onster,'' in {\em Vertex operators in mathematics and physics (Berkeley,
  Calif., 1983)}, vol.~3 of {\em Math. Sci. Res. Inst. Publ.}, pp.~231--273.
\newblock Springer, New York, 1985.

\bibitem{borcherds_monstrous}
R.~E. Borcherds, ``{Monstrous moonshine and monstrous Lie superalgebras},''
  {\em Invent. Math.} {\bf 109, No.2} (1992)  405--444.

\bibitem{MR2385372}
J.~H. Bruinier, G.~van~der Geer, G.~Harder, and D.~Zagier, {\em The 1-2-3 of
  modular forms}.
\newblock Universitext. Springer-Verlag, Berlin, 2008.
\newblock Lectures from the Summer School on Modular Forms and their
  Applications held in Nordfjordeid, June 2004, Edited by Kristian Ranestad.

\bibitem{gannon}
T.~Gannon, {\em {Moonshine beyond the monster. The bridge connecting algebra,
  modular forms and physics}}.
\newblock Cambridge University Press, 2006.

\bibitem{UM}
M.~C.~N. Cheng, J.~F.~R. Duncan, and J.~A. Harvey, ``{Umbral Moonshine},''
\href{http://arxiv.org/abs/1204.2779}{{\tt arXiv:1204.2779 [math.RT]}}.

\bibitem{Folsom_what}
A.~Folsom, ``What is {$\dots$} a mock modular form?,'' {\em Notices Amer. Math.
  Soc.} {\bf 57} (2010) no.~11, 1441--1443.

\bibitem{zagier_mock}
D.~Zagier, ``Ramanujan's mock theta functions and their applications (after
  {Z}wegers and {O}no-{B}ringmann),'' {\em Ast\'erisque} (2009) no.~326, Exp.
  No. 986, vii--viii, 143--164 (2010). S{\'e}minaire Bourbaki. Vol. 2007/2008.

\bibitem{VafaWitten1994}
C.~Vafa and E.~Witten, ``A strong coupling test of {$S$}-duality,''
  \href{http://dx.doi.org/10.1016/0550-3213(94)90097-3}{{\em Nuclear Phys. B}
  {\bf 431} (1994) no.~1-2, 3--77}.
  \url{http://dx.doi.org/10.1016/0550-3213(94)90097-3}.

\bibitem{Troost:2010ud}
J.~Troost, ``{The non-compact elliptic genus: mock or modular},''
  \href{http://dx.doi.org/10.1007/JHEP06(2010)104}{{\em JHEP} {\bf 1006} (2010)
   104},
\href{http://arxiv.org/abs/1004.3649}{{\tt arXiv:1004.3649 [hep-th]}}.

\bibitem{Dabholkar:2012nd}
A.~Dabholkar, S.~Murthy, and D.~Zagier, ``{Quantum Black Holes, Wall Crossing,
  and Mock Modular Forms},''
\href{http://arxiv.org/abs/1208.4074}{{\tt arXiv:1208.4074 [hep-th]}}.

\bibitem{Alexandrov:2012au}
S.~Alexandrov, J.~Manschot, and B.~Pioline, ``{D3-instantons, Mock Theta Series
  and Twistors},'' \href{http://dx.doi.org/10.1007/JHEP04(2013)002}{{\em JHEP}
  {\bf 1304} (2013)  002},
\href{http://arxiv.org/abs/1207.1109}{{\tt arXiv:1207.1109 [hep-th]}}.

\bibitem{Eguchi1987}
T.~Eguchi and A.~Taormina, ``Unitary representations of the {$N=4$}
  superconformal algebra,''
  \href{http://dx.doi.org/10.1016/0370-2693(87)91679-0}{{\em Phys. Lett. B}
  {\bf 196} (1987) no.~1, 75--81}.
  \url{http://dx.doi.org/10.1016/0370-2693(87)91679-0}.

\bibitem{Eguchi1988}
T.~Eguchi and A.~Taormina, ``Character formulas for the {$N=4$} superconformal
  algebra,'' \href{http://dx.doi.org/10.1016/0370-2693(88)90778-2}{{\em Phys.
  Lett. B} {\bf 200} (1988) no.~3, 315--322}.
  \url{http://dx.doi.org/10.1016/0370-2693(88)90778-2}.

\bibitem{Eguchi1989}
T.~Eguchi, H.~Ooguri, A.~Taormina, and S.-K. Yang, ``{Superconformal Algebras
  and String Compactification on Manifolds with SU(N) Holonomy},''
\href{http://dx.doi.org/10.1016/0550-3213(89)90454-9}{{\em Nucl. Phys.} {\bf
  B315} (1989)  193}.

\bibitem{Eguchi2010}
T.~Eguchi, H.~Ooguri, and Y.~Tachikawa, ``{Notes on the K3 Surface and the
  Mathieu group $M_{24}$},'' {\em Exper.Math.} {\bf 20} (2011)  91--96,
\href{http://arxiv.org/abs/1004.0956}{{\tt arXiv:1004.0956 [hep-th]}}.

\bibitem{Cheng2010_1}
M.~C. Cheng, ``{K3 Surfaces, N=4 Dyons, and the Mathieu Group {$M_{24}$}},''
  \href{http://dx.doi.org/10.4310/CNTP.2010.v4.n4.a2}{{\em
  Commun.Num.Theor.Phys.} {\bf 4} (2010)  623--658},
\href{http://arxiv.org/abs/1005.5415}{{\tt arXiv:1005.5415 [hep-th]}}.

\bibitem{Gaberdiel2010}
M.~R. Gaberdiel, S.~Hohenegger, and R.~Volpato, ``{Mathieu twining characters
  for K3},'' \href{http://dx.doi.org/10.1007/JHEP09(2010)058}{{\em JHEP} {\bf
  1009} (2010)  058},
\href{http://arxiv.org/abs/1006.0221}{{\tt arXiv:1006.0221 [hep-th]}}.

\bibitem{Gaberdiel2010a}
M.~R. Gaberdiel, S.~Hohenegger, and R.~Volpato, ``{Mathieu Moonshine in the
  elliptic genus of K3},''
  \href{http://dx.doi.org/10.1007/JHEP10(2010)062}{{\em JHEP} {\bf 1010} (2010)
   062},
\href{http://arxiv.org/abs/1008.3778}{{\tt arXiv:1008.3778 [hep-th]}}.

\bibitem{Eguchi2010a}
T.~Eguchi and K.~Hikami, ``{Note on Twisted Elliptic Genus of K3 Surface},''
  \href{http://dx.doi.org/10.1016/j.physletb.2010.10.017}{{\em Phys.Lett.} {\bf
  B694} (2011)  446--455},
\href{http://arxiv.org/abs/1008.4924}{{\tt arXiv:1008.4924 [hep-th]}}.

\bibitem{Cheng2011}
M.~C. Cheng and J.~F. Duncan, ``{On Rademacher Sums, the Largest Mathieu Group,
  and the Holographic Modularity of Moonshine},''
  \href{http://dx.doi.org/10.4310/CNTP.2012.v6.n3.a4}{{\em
  Commun.Num.Theor.Phys.} {\bf 6} (2012)  697--758},
\href{http://arxiv.org/abs/1110.3859}{{\tt arXiv:1110.3859 [math.RT]}}.

\bibitem{Gaberdiel:2012gf}
M.~R. Gaberdiel, D.~Persson, H.~Ronellenfitsch, and R.~Volpato, ``{Generalised
  Mathieu Moonshine},''
  \href{http://dx.doi.org/10.4310/CNTP.2013.v7.n1.a5}{{\em
  Commun.Num.Theor.Phys.} {\bf 7} (2013)  145--223},
\href{http://arxiv.org/abs/1211.7074}{{\tt arXiv:1211.7074 [hep-th]}}.

\bibitem{2012arXiv1212.0906C}
M.~C.~N. {Cheng} and J.~F.~R. {Duncan}, ``{On the Discrete Groups of Mathieu
  Moonshine},''{\em ArXiv e-prints} (Dec., 2012)  ,
  \href{http://arxiv.org/abs/1212.0906}{{\tt arXiv:1212.0906 [math.NT]}}.

\bibitem{Gaberdiel:2013nya}
M.~R. Gaberdiel, D.~Persson, and R.~Volpato, ``{Generalised Moonshine and
  Holomorphic Orbifolds},''
\href{http://arxiv.org/abs/1302.5425}{{\tt arXiv:1302.5425 [hep-th]}}.

\bibitem{Persson:2013xpa}
D.~Persson and R.~Volpato, ``{Second Quantized Mathieu Moonshine},''
\href{http://arxiv.org/abs/1312.0622}{{\tt arXiv:1312.0622 [hep-th]}}.

\bibitem{Raum}
M.~Raum, ``M24-twisted product expansions are siegel modular forms,''
  \href{http://arxiv.org/abs/1208.3453}{{\tt arXiv:1208.3453 [math.NT]}}.

\bibitem{Gannon:2012ck}
T.~Gannon, ``{Much ado about Mathieu},''
\href{http://arxiv.org/abs/1211.5531}{{\tt arXiv:1211.5531 [math.RT]}}.

\bibitem{CheDun_M24MckAutFrms}
M.~C.~N. Cheng and J.~F.~R. Duncan, ``{The Largest Mathieu Group and (Mock)
  Automorphic Forms},'' \href{http://arxiv.org/abs/1201.4140}{{\tt 1201.4140
  [math.RT]}}.

\bibitem{Taormina2010}
A.~Taormina and K.~Wendland, ``The symmetries of the tetrahedral kummer surface
  in the mathieu group {$M_{24}$},'' \href{http://arxiv.org/abs/1008.0954}{{\tt
  1008.0954}}.

\bibitem{Taormina:2011rr}
A.~Taormina and K.~Wendland, ``{The overarching finite symmetry group of Kummer
  surfaces in the Mathieu group $M_{24}$},''
  \href{http://dx.doi.org/10.1007/JHEP08(2013)125}{{\em JHEP} {\bf 1308} (2013)
   125},
\href{http://arxiv.org/abs/1107.3834}{{\tt arXiv:1107.3834 [hep-th]}}.

\bibitem{Gaberdiel2011}
M.~R. Gaberdiel, S.~Hohenegger, and R.~Volpato, ``{Symmetries of K3 sigma
  models},'' \href{http://dx.doi.org/10.4310/CNTP.2012.v6.n1.a1}{{\em
  Commun.Num.Theor.Phys.} {\bf 6} (2012)  1--50},
\href{http://arxiv.org/abs/1106.4315}{{\tt arXiv:1106.4315 [hep-th]}}.

\bibitem{Govindarajan:2011em}
S.~Govindarajan, ``{Unravelling Mathieu Moonshine},''
  \href{http://dx.doi.org/10.1016/j.nuclphysb.2012.07.005}{{\em Nucl.Phys.}
  {\bf B864} (2012)  823--839},
\href{http://arxiv.org/abs/1106.5715}{{\tt arXiv:1106.5715 [hep-th]}}.

\bibitem{Cheng:2013kpa}
M.~C. Cheng, X.~Dong, J.~Duncan, J.~Harvey, S.~Kachru, {\em et al.}, ``{Mathieu
  Moonshine and N=2 String Compactifications},''
  \href{http://dx.doi.org/10.1007/JHEP09(2013)030}{{\em JHEP} {\bf 1309} (2013)
   030},
\href{http://arxiv.org/abs/1306.4981}{{\tt arXiv:1306.4981 [hep-th]}}.

\bibitem{Taormina:2013jza}
A.~Taormina and K.~Wendland, ``{Symmetry-surfing the moduli space of Kummer
  K3s},''
\href{http://arxiv.org/abs/1303.2931}{{\tt arXiv:1303.2931 [hep-th]}}.

\bibitem{Harrison:2013bya}
S.~Harrison, S.~Kachru, and N.~M. Paquette, ``{Twining Genera of (0,4)
  Supersymmetric Sigma Models on K3},''
  \href{http://dx.doi.org/10.1007/JHEP04(2014)048}{{\em JHEP} {\bf 1404} (2014)
   048},
\href{http://arxiv.org/abs/1309.0510}{{\tt arXiv:1309.0510 [hep-th]}}.

\bibitem{Harvey:2013mda}
J.~A. Harvey and S.~Murthy, ``{Moonshine in Fivebrane Spacetimes},''
  \href{http://dx.doi.org/10.1007/JHEP01(2014)146}{{\em JHEP} {\bf 1401} (2014)
   146},
\href{http://arxiv.org/abs/1307.7717}{{\tt arXiv:1307.7717 [hep-th]}}.

\bibitem{Wrase:2014fja}
T.~Wrase, ``{Mathieu moonshine in four dimensional $\mathcal{N}=1$ theories},''
  \href{http://dx.doi.org/10.1007/JHEP04(2014)069}{{\em JHEP} {\bf 1404} (2014)
   069},
\href{http://arxiv.org/abs/1402.2973}{{\tt arXiv:1402.2973 [hep-th]}}.

\bibitem{Creutzig:2013mqa}
T.~Creutzig and G.~Hoehn, ``{Mathieu Moonshine and the Geometry of K3
  Surfaces},''
\href{http://arxiv.org/abs/1309.2671}{{\tt arXiv:1309.2671 [math.QA]}}.

\bibitem{John_Sander}
J.~Duncan and S.~Mack-Crane, ``{Derived Equivalences of K3 Surfaces and Twined
  Ellipitic Genera},'' {\em to appear}  .

\bibitem{Cheng:2013wca}
M.~C.~N. Cheng, J.~F.~R. Duncan, and J.~A. Harvey, ``{Umbral Moonshine and the
  Niemeier Lattices},''
\href{http://arxiv.org/abs/1307.5793}{{\tt arXiv:1307.5793 [math.RT]}}.

\bibitem{Cappelli:1987xt}
A.~Cappelli, C.~Itzykson, and J.~Zuber, ``{The ADE Classification of Minimal
  and A1(1) Conformal Invariant Theories},''
\href{http://dx.doi.org/10.1007/BF01221394}{{\em Commun.Math.Phys.} {\bf 113}
  (1987)  1}.

\bibitem{Cappelli:1986hf}
A.~Cappelli, C.~Itzykson, and J.~Zuber, ``{Modular Invariant Partition
  Functions in Two-Dimensions},''
\href{http://dx.doi.org/10.1016/0550-3213(87)90155-6}{{\em Nucl.Phys.} {\bf
  B280} (1987)  445--465}.

\bibitem{Gepner:1987qi}
D.~Gepner, ``{Space-Time Supersymmetry in Compactified String Theory and
  Superconformal Models},''
\href{http://dx.doi.org/10.1016/0550-3213(88)90397-5}{{\em Nucl.Phys.} {\bf
  B296} (1988)  757}.

\bibitem{Vafa:1988uu}
C.~Vafa and N.~P. Warner, ``{Catastrophes and the Classification of Conformal
  Theories},''
\href{http://dx.doi.org/10.1016/0370-2693(89)90473-5}{{\em Phys.Lett.} {\bf
  B218} (1989)  51}.

\bibitem{Martinec:1988zu}
E.~J. Martinec, ``{Algebraic Geometry and Effective Lagrangians},''
\href{http://dx.doi.org/10.1016/0370-2693(89)90074-9}{{\em Phys.Lett.} {\bf
  B217} (1989)  431}.

\bibitem{MR0084174}
H.~Cartan, ``Quotient d'un espace analytique par un groupe d'automorphismes,''
  in {\em Algebraic geometry and topology}, pp.~90--102.
\newblock Princeton University Press, Princeton, N. J., 1957.
\newblock A symposium in honor of S. Lefschetz,.

\bibitem{Nikulin:2011}
V.~V. Nikulin, ``{K\"ahlerian K3 Surfaces and Niemeier Lattices},''
  \href{http://arxiv.org/abs/1109.2879}{{\tt arXiv:1109.2879 [math.AG]}}.

\bibitem{Nikulin:2014}
V.~V. Nikulin, ``{Degenerations of K\"ahlerian K3 surfaces with finite
  symplectic automorphism groups },''
  \href{http://arxiv.org/abs/1403.6061}{{\tt arXiv:1403.6061 [math.AG]}}.

\bibitem{MR543555}
A.~H. Durfee, ``Fifteen characterizations of rational double points and simple
  critical points,'' {\em Enseign. Math. (2)} {\bf 25} (1979) no.~1-2,
  131--163.

\bibitem{witten_string_1995}
E.~Witten, \href{http://dx.doi.org/10.1016/0550-3213(95)00158-O}{``String
  theory dynamics in various dimensions,''{\em Nuclear Physics B} {\bf 443}
  (June, 1995)  }. \url{http://arxiv.org/abs/hep-th/9503124}.

\bibitem{Aspinwall:1995zi}
P.~S. Aspinwall, ``{Enhanced gauge symmetries and K3 surfaces},''
  \href{http://dx.doi.org/10.1016/0370-2693(95)00957-M}{{\em Phys.Lett.} {\bf
  B357} (1995)  329--334},
\href{http://arxiv.org/abs/hep-th/9507012}{{\tt arXiv:hep-th/9507012
  [hep-th]}}.

\bibitem{Katz:1996fh}
S.~H. Katz, A.~Klemm, and C.~Vafa, ``{Geometric engineering of quantum field
  theories},'' \href{http://dx.doi.org/10.1016/S0550-3213(97)00282-4}{{\em
  Nucl.Phys.} {\bf B497} (1997)  173--195},
\href{http://arxiv.org/abs/hep-th/9609239}{{\tt arXiv:hep-th/9609239
  [hep-th]}}.

\bibitem{ooguri_two-dimensional_1995}
H.~Ooguri and C.~Vafa, ``Two-dimensional black hole and singularities of {CY}
  manifolds,''. \url{http://arxiv.org/abs/hep-th/9511164}.
  {Nucl.Phys.B463:55-72},1996.

\bibitem{Kazama:1988qp}
Y.~Kazama and H.~Suzuki, ``{New N=2 Superconformal Field Theories and
  Superstring Compactification},''
\href{http://dx.doi.org/10.1016/0550-3213(89)90250-2}{{\em Nucl.Phys.} {\bf
  B321} (1989)  232}.

\bibitem{Giveon:1999px}
A.~Giveon and D.~Kutasov, ``{Little string theory in a double scaling limit},''
  \href{http://dx.doi.org/10.1088/1126-6708/1999/10/034}{{\em JHEP} {\bf 9910}
  (1999)  034},
\href{http://arxiv.org/abs/hep-th/9909110}{{\tt arXiv:hep-th/9909110
  [hep-th]}}.

\bibitem{Harvey:2014cva}
J.~A. Harvey, S.~Murthy, and C.~Nazaroglu, ``{ADE Double Scaled Little String
  Theories, Mock Modular Forms and Umbral Moonshine},''
\href{http://arxiv.org/abs/1410.6174}{{\tt arXiv:1410.6174 [hep-th]}}.

\bibitem{MR1449324}
T.~Gannon, ``{${\rm U}(1)^m$} modular invariants, {$N=2$} minimal models, and
  the quantum {H}all effect,''
  \href{http://dx.doi.org/10.1016/S0550-3213(97)00032-1}{{\em Nuclear Phys. B}
  {\bf 491} (1997) no.~3, 659--688}.
  \url{http://dx.doi.org/10.1016/S0550-3213(97)00032-1}.

\bibitem{MR2925130}
O.~Gray, ``On the complete classification of unitary {$N=2$} minimal
  superconformal field theories,''
  \href{http://dx.doi.org/10.1007/s00220-012-1478-z}{{\em Comm. Math. Phys.}
  {\bf 312} (2012) no.~3, 611--654}.
  \url{http://dx.doi.org/10.1007/s00220-012-1478-z}.

\bibitem{Witten:1993yc}
E.~Witten, ``{Phases of N=2 theories in two-dimensions},''
  \href{http://dx.doi.org/10.1016/0550-3213(93)90033-L}{{\em Nucl.Phys.} {\bf
  B403} (1993)  159--222},
\href{http://arxiv.org/abs/hep-th/9301042}{{\tt arXiv:hep-th/9301042
  [hep-th]}}.

\bibitem{WittenInt.J.Mod.Phys.A9:4783-48001994}
E.~Witten, ``{On the Landau-Ginzburg Description of $N=2$ Minimal Models},''
  {\em Int.J.Mod.Phys.A} {\bf 9:4783-4800,1994}
  (Int.J.Mod.Phys.A9:4783-4800,1994)  ,
  \href{http://arxiv.org/abs/hep-th/9304026}{{\tt hep-th/9304026}}.

\bibitem{Qiu:1987ux}
Z.-a. Qiu, ``{Modular Invariant Partition Functions for $N=2$ Superconformal
  Field Theories},''
\href{http://dx.doi.org/10.1016/0370-2693(87)90906-3}{{\em Phys.Lett.} {\bf
  B198} (1987)  497}.

\bibitem{kawai_elliptic_1993}
T.~Kawai, Y.~Yamada, and S.-K. Yang, ``Elliptic {G}enera and {N}=2
  {S}uperconformal {F}ield {T}heory,''.
  \url{http://arxiv.org/abs/hep-th/9306096}. {Nucl.Phys.B414:191-212},1994.

\bibitem{DiFrancesco:1993dg}
P.~Di~Francesco and S.~Yankielowicz, ``{Ramond sector characters and N=2
  Landau-Ginzburg models},''
  \href{http://dx.doi.org/10.1016/0550-3213(93)90452-U}{{\em Nucl.Phys.} {\bf
  B409} (1993)  186--210},
\href{http://arxiv.org/abs/hep-th/9305037}{{\tt arXiv:hep-th/9305037
  [hep-th]}}.

\bibitem{Witten:1991yr}
E.~Witten, ``{On string theory and black holes},''
\href{http://dx.doi.org/10.1103/PhysRevD.44.314}{{\em Phys.Rev.} {\bf D44}
  (1991)  314--324}.

\bibitem{Hori:2001ax}
K.~Hori and A.~Kapustin, ``{Duality of the fermionic 2-D black hole and N=2
  liouville theory as mirror symmetry},''
  \href{http://dx.doi.org/10.1088/1126-6708/2001/08/045}{{\em JHEP} {\bf 0108}
  (2001)  045},
\href{http://arxiv.org/abs/hep-th/0104202}{{\tt arXiv:hep-th/0104202
  [hep-th]}}.

\bibitem{Israel:2004ir}
D.~Israel, C.~Kounnas, A.~Pakman, and J.~Troost, ``{The Partition function of
  the supersymmetric two-dimensional black hole and little string theory},''
  \href{http://dx.doi.org/10.1088/1126-6708/2004/06/033}{{\em JHEP} {\bf 0406}
  (2004)  033},
\href{http://arxiv.org/abs/hep-th/0403237}{{\tt arXiv:hep-th/0403237
  [hep-th]}}.

\bibitem{Eguchi:2004yi}
T.~Eguchi and Y.~Sugawara, ``{SL(2,R)/U(1) supercoset and elliptic genera of
  noncompact Calabi-Yau manifolds},''
  \href{http://dx.doi.org/10.1088/1126-6708/2004/05/014}{{\em JHEP} {\bf 0405}
  (2004)  014},
\href{http://arxiv.org/abs/hep-th/0403193}{{\tt arXiv:hep-th/0403193
  [hep-th]}}.

\bibitem{Dixon:1989cg}
L.~J. Dixon, M.~E. Peskin, and J.~D. Lykken, ``{N=2 Superconformal Symmetry and
  SO(2,1) Current Algebra},''
\href{http://dx.doi.org/10.1016/0550-3213(89)90459-8}{{\em Nucl.Phys.} {\bf
  B325} (1989)  329--355}.

\bibitem{Hanany:2002ev}
A.~Hanany, N.~Prezas, and J.~Troost, ``{The Partition function of the
  two-dimensional black hole conformal field theory},''
  \href{http://dx.doi.org/10.1088/1126-6708/2002/04/014}{{\em JHEP} {\bf 0204}
  (2002)  014},
\href{http://arxiv.org/abs/hep-th/0202129}{{\tt arXiv:hep-th/0202129
  [hep-th]}}.

\bibitem{Dijkgraaf:1991ba}
R.~Dijkgraaf, H.~L. Verlinde, and E.~P. Verlinde, ``{String propagation in a
  black hole geometry},''
\href{http://dx.doi.org/10.1016/0550-3213(92)90237-6}{{\em Nucl.Phys.} {\bf
  B371} (1992)  269--314}.

\bibitem{Eguchi:2010cb}
T.~Eguchi and Y.~Sugawara, ``{Non-holomorphic Modular Forms and SL(2,R)/U(1)
  Superconformal Field Theory},''
  \href{http://dx.doi.org/10.1007/JHEP03(2011)107}{{\em JHEP} {\bf 1103} (2011)
   107},
\href{http://arxiv.org/abs/1012.5721}{{\tt arXiv:1012.5721 [hep-th]}}.

\bibitem{Ashok:2011cy}
S.~K. Ashok and J.~Troost, ``{A Twisted Non-compact Elliptic Genus},''
  \href{http://dx.doi.org/10.1007/JHEP03(2011)067}{{\em JHEP} {\bf 1103} (2011)
   067},
\href{http://arxiv.org/abs/1101.1059}{{\tt arXiv:1101.1059 [hep-th]}}.

\bibitem{Zwegers2008}
S.~Zwegers, ``Mock theta functions,''
  \href{http://arxiv.org/abs/0807.4834}{{\tt arXiv:0807.4834 [math.RT]}}.

\bibitem{Ashok:2013pya}
S.~K. Ashok, N.~Doroud, and J.~Troost, ``{Localization and real Jacobi
  forms},''
\href{http://arxiv.org/abs/1311.1110}{{\tt arXiv:1311.1110 [hep-th]}}.

\bibitem{Murthy:2013mya}
S.~Murthy, ``{A holomorphic anomaly in the elliptic genus},''
\href{http://arxiv.org/abs/1311.0918}{{\tt arXiv:1311.0918 [hep-th]}}.

\bibitem{Ochanine}
S.~Ochanine {\em Topology. An International Journal of Mathematics} {\bf 26}
  (1987)  143.

\bibitem{Witten1987}
E.~Witten, ``{ELLIPTIC GENERA AND QUANTUM FIELD THEORY},''
\href{http://dx.doi.org/10.1007/BF01208956}{{\em Commun. Math. Phys.} {\bf 109}
  (1987)  525}.

\bibitem{Landweber_book}
P.~S. Landweber, ed., {\em {Elliptic curves and modular forms in algebraic
  topology}}.
\newblock Springer-Verlag, Berlin, 1988.

\bibitem{KawaiNucl.Phys.B414:191-2121994}
T.~Kawai, Y.~Yamada, and S.-K. Yang, ``Elliptic genera and n=2 superconformal
  field theory,'' {\em Nucl.Phys.B} {\bf 414:191-212,1994}
  (Nucl.Phys.B414:191-212,1994)  ,
  \href{http://arxiv.org/abs/hep-th/9306096}{{\tt hep-th/9306096}}.

\bibitem{Kapustin:2005pt}
A.~Kapustin, ``{Chiral de Rham complex and the half-twisted sigma-model},''
\href{http://arxiv.org/abs/hep-th/0504074}{{\tt arXiv:hep-th/0504074
  [hep-th]}}.

\bibitem{Nie_DefQdtFrm24}
H.-V. Niemeier, ``Definite quadratische {F}ormen der {D}imension {$24$} und
  {D}iskriminante {$1$},'' {\em J. Number Theory} {\bf 5} (1973)  142--178.

\bibitem{Con_ChrLeeLat}
J.~H. Conway, ``A characterisation of {L}eech's lattice,'' {\em Invent. Math.}
  {\bf 7} (1969)  137--142.

\bibitem{Lee_SphPkgs}
J.~Leech, ``Notes on sphere packings,'' {\em Canad. J. Math.} {\bf 19} (1967)
  251--267.

\bibitem{Lee_SphPkgHgrSpc}
J.~Leech, ``Some sphere packings in higher space,'' {\em Canad. J. Math.} {\bf
  16} (1964)  657--682.

\bibitem{Sko_Thesis}
N.-P. Skoruppa, {\em {\"U}ber den {Z}usammenhang zwischen {J}acobiformen und
  {M}odulformen halbganzen {G}ewichts}.
\newblock Bonner Mathematische Schriften [Bonn Mathematical Publications], 159.
  Universit{\"a}t Bonn Mathematisches Institut, Bonn, 1985.
\newblock Dissertation, Rheinische Friedrich-Wilhelms-Universit{\"a}t, Bonn,
  1984.

\bibitem{Eguchi1988a}
T.~Eguchi and A.~Taormina, ``On the unitary representations of {$N=2$} and
  {$N=4$} superconformal algebras,''
  \href{http://dx.doi.org/10.1016/0370-2693(88)90360-7}{{\em Phys. Lett. B}
  {\bf 210} (1988) no.~1-2, 125--132}.
  \url{http://dx.doi.org/10.1016/0370-2693(88)90360-7}.

\bibitem{Kondo}
S.~Kond{\=o}, ``Niemeier lattices, {M}athieu groups, and finite groups of
  symplectic automorphisms of {$K3$} surfaces,''
  \href{http://dx.doi.org/10.1215/S0012-7094-98-09217-1}{{\em Duke Math. J.}
  {\bf 92} (1998) no.~3, 593--603}.
  \url{http://dx.doi.org/10.1215/S0012-7094-98-09217-1}. With an appendix by
  Shigeru Mukai.

\bibitem{Nikulin1980}
V.~V. Nikulin, ``Integral symmetric bilinear forms and some of their
  applications,'' {\em Mathematics of the USSR-Izvestiya} {\bf 14} (1980)
  no.~1, 103. \url{http://stacks.iop.org/0025-5726/14/i=1/a=A06}.

\bibitem{MR0409904}
V.~V. Nikulin, ``Finite groups of automorphisms of {K}\"ahlerian surfaces of
  type {$K3$},'' {\em Uspehi Mat. Nauk} {\bf 31} (1976) no.~2(188), 223--224.

\bibitem{Mukai}
S.~Mukai, ``Finite groups of automorphisms of {$K3$} surfaces and the {M}athieu
  group,'' \href{http://dx.doi.org/10.1007/BF01394352}{{\em Invent. Math.} {\bf
  94} (1988) no.~1, 183--221}. \url{http://dx.doi.org/10.1007/BF01394352}.

\bibitem{MR544937}
V.~V. Nikulin, ``Finite groups of automorphisms of {K}\"ahlerian {$K3$}
  surfaces,'' {\em Trudy Moskov. Mat. Obshch.} {\bf 38} (1979)  75--137.

\bibitem{MR2926486}
K.~Hashimoto, ``Finite symplectic actions on the {$K3$} lattice,'' {\em Nagoya
  Math. J.} {\bf 206} (2012)  99--153.
  \url{http://projecteuclid.org/euclid.nmj/1337690053}.

\bibitem{Nik_Kummer}
V.~V. Nikulin, ``On {K}ummer {S}urfaces,'' {\em Izv. Akad. Nauk SSSR Ser. Mat.}
   .

\bibitem{Fujiki}
A.~Fujiki, ``{Finite automorphism groups of complex tori of dimension 2},''
  {\em Publ. RIMS. Kyoto Univ.} {\bf 24} (1988)  1--97.

\bibitem{Wendland:2000ry}
K.~Wendland, ``{Consistency of orbifold conformal field theories on K3},'' {\em
  Adv.Theor.Math.Phys.} {\bf 5} (2002)  429--456,
\href{http://arxiv.org/abs/hep-th/0010281}{{\tt arXiv:hep-th/0010281
  [hep-th]}}.

\bibitem{Aspinwall:1994rg}
P.~S. Aspinwall and D.~R. Morrison, ``{String theory on K3 surfaces},''
\href{http://arxiv.org/abs/hep-th/9404151}{{\tt arXiv:hep-th/9404151
  [hep-th]}}.

\bibitem{eichler_zagier}
M.~Eichler and D.~Zagier, {\em {The theory of Jacobi forms}}.
\newblock Birkh{\"a}user, 1985.

\end{thebibliography}\endgroup

\end{document}